\documentclass[apj]{emulateapj}
\usepackage{bm}
\usepackage{multirow}
\usepackage{color}
\bibpunct[; ]{(}{)}{;}{a}{}{;}

\usepackage{amsmath}
\usepackage{amsfonts}
\usepackage{amssymb}
\usepackage{epsfig}
\usepackage{hyperref}
\usepackage{graphicx}

\newcommand{\beq}{\begin{equation}}
\newcommand{\eeq}{\end{equation}}
\newcommand{\beqa}{\begin{eqnarray}}
\newcommand{\eeqa}{\end{eqnarray}}

\begin{document}

\title{H-ATLAS: The Cosmic Abundance of Dust from the Far-Infrared Background Power Spectrum}
\author{ Cameron Thacker$^{1}$, Asantha Cooray$^1$, Joseph Smidt$^{1}$, Francesco De
Bernardis$^{1}$, K. Mitchell-Wynne$^{1}$, A. Amblard$^2$, R. Auld$^3$, M. Baes$^{4}$,
D. L. Clements$^5$, A. Dariush$^5$, G. De Zotti$^6$, L. Dunne$^7$, S. Eales$^3$, R. Hopwood$^5$, 
C. Hoyos$^8$, E. Ibar$^{9,10}$, M. Jarvis$^{11,12}$, S. Maddox$^7$, M. J. Micha{\l}owski$^{4,13,14}$,
E. Pascale$^3$, D. Scott$^{15}$, S. Serjeant$^{16}$, M. W. L. Smith$^3$, E. Valiante$^3$,
P. van der Werf$^{17}$}

\affiliation{$^{1}$Department of Physics  and  Astronomy, University  of California, Irvine, CA 92697}
\affiliation{$^{2}$NASA Ames Research Center, Moffett Field, CA 94035}
\affiliation{$^{3}$School of Physics and Astronomy, Cardiff University, The Parade, Cardiff, CF24 3AA, UK}
\affiliation{$^{4}$Sterrenkundig Observatorium, Universiteit Gent, KrijgslAAn 281 S9, B-9000 Gent, Belgium }
\affiliation{$^5$Physics Department, Imperial College London, South Kensington campus, London, SW7 2AZ, UK}
\affiliation{$^{6}$INAF, Osservatorio Astronomico di Padova, Vicolo Osservatorio 5, I-35122 Padova, Italy}
\affiliation{$^{7}$Department of Physics and Astronomy, University of Canterbury, Private Bag 4800, Christchurch, New Zealand}
\affiliation{$^{8}$School of Physics and Astronomy, University of Nottingham, University Park, Nottingham NG7 2RD, UK}
\affiliation{$^{9}$UK Astronomy Technology Centre, The Royal Observatory, Blackford Hill, Edinburgh, EH9 3HJ, UK}
\affiliation{$^{10}$Universidad Cat\'olica de Chile, Departamento de Astronom\'ia y Astrof\'isica, Vicu\~na Mackenna 4860, Casilla 306, Santiago 22, Chile }
\affiliation{$^{11}$Astrophysics, Department of Physics, Keble Road, Oxford OX1 3RH, UK}
\affiliation{$^{12}$Department of Physics, University of the Western Cape, Private Bag X17, Bellville 7535, South Africa}
\affiliation{$^{13}$Scottish Universities Physics Alliance, Institute for Astronomy, University of Edinburgh, Royal Observatory, Edinburgh, EH9 3HJ, UK}
\affiliation{$^{14}$FWO Pegasus Marie Curie Fellow}
\affiliation{$^{15}$Department of Physics \& Astronomy, University of British Columbia, 6224 Agricultural Road, Vancouver, BC V6T 1Z1, Canada}
\affiliation{$^{16}$Department of Physical Sciences, The Open University, Milton Keynes MK7 6AA, UK}
\affiliation{$^{17}$Leiden Observatory, Leiden University, P.O. Box 9513, NL-2300 RA Leiden, The Netherlands}

\begin{abstract}
We present a measurement of the angular power spectrum of the cosmic far-infrared background (CFIRB) anisotropies in one of the extragalactic fields of the
{\it Herschel} Astrophysical Terahertz Large Area Survey (H-ATLAS) at 250, 350 and 500\,\micron\ bands. Consistent with recent measurements of the CFIRB power spectrum in
{\it Herschel}-SPIRE maps,  we confirm the existence of a clear one-halo term of galaxy clustering on arcminute angular scales with large-scale two-halo term of
clustering at 30 arcminutes to angular scales of a few degrees. The power spectrum at the largest
angular scales, especially at 250\,$\mu$m, is contaminated by the Galactic cirrus.
The angular power spectrum is modeled using a conditional luminosity function approach to describe the spatial distribution of unresolved galaxies that make up the
bulk of the CFIRB.  Integrating over the dusty galaxy population responsible for the background anisotropies, we find that the cosmic abundance of dust,
relative to the critical density, to be  between $\Omega_{\rm dust}=10^{-6}$ and $8\times 10^{-6}$ in the redshift range $z \sim 0-3$.
This dust abundance is consistent with estimates of the dust content in the Universe using quasar reddening and magnification measurements in the SDSS.
\end{abstract}

\maketitle

\section{Introduction}

While the total intensity of the cosmic far infrared background (CFIRB)
is known from absolute photometry measurements \citep{Puget1996,Fixen1998, Dwek1998}, we still lack a complete knowledge of the sources, in the form of dusty star-forming galaxies,
 that make up the background. Limited by aperture sizes and the resulting source confusion noise \citep{Nguyen2010}, existing deep surveys with
the {\it Herschel} Space Observatory \footnote{{\it Herschel} is an ESA space observatory with science
instruments provided by European-led Principal Investigator consortia and with important
participation from NASA.} (Pilbratt et al. 2010) and ground-based sub-mm and mm-wave instruments
resolve anywhere between 5 and 15\% of
the background into individual galaxies \citep{Coppin2006,Scott:2010dv,hermes,Clements2010,Berta2011}.  Anisotropies of the CFIRB, or the spatial fluctuations of the background intensity,
provide additional statistical information on the fainter galaxies, especially those that make up the bulk of the background.

While the fainter galaxies are individually undetected, due to gravitational growth and evolution in the large-scale structure these galaxies
are expected to be clustered \citep{Cooray2010, Maddox2010, Hickox2012, Kampen2012}. In the ansatz of the halo model
\citep{Cooray2002} such clustering of galaxies captures certain  properties of the
dark matter halos in which galaxies are found and the statistics of how those galaxies occupy the dark matter halos. The resulting anisotropies of the CFIRB
are then a reflection of the spatial clustering of galaxies and their infrared luminosity.
These CFIRB anisotropies, are best studied from the angular power spectrum of the background infrared light. Separately,
statistics such as the probability of deflection, $P(D)$ \citep{Glenn2010}, probe the variance and higher order cumulant statistics of the intensity variations at the beam scale.

While early attempts to measure the angular power spectrum of the CFIRB resulted in low signal-to-noise measurements \citep{Lagache2007, Viero2009},
a first clear detection of the CFIRB power spectrum with {\it Herschel}-SPIRE (Griffin et al. 2010) 
maps between 30 arcseconds and 30 arcminute angular scales was reported in \citet{Amblard:2011gc}.
Those first measurements also confirmed the interpretation that galaxies at the peak epoch of star formation in the Universe at redshifts of 1 to 3 trace the
underlying dark matter halo distribution. Since then, additional measurements of the CFIRB power spectrum have come from Planck \citep{Planck2011}
and with additional SPIRE maps from the HerMES survey \citep{Viero2012}. With multiple fields
spanning up to 20 deg$^2$, recent HerMES CFIRB power spectra
probe angular scales of about 30\arcsec\: to 2$^{\circ}$. The halo model interpretation of the
HerMES spectra suggest that the halo mass scale for peak star-formation
activity is $\log M_{\rm peak}/M_{\sun} \sim 13.9 \pm 0.6$ and the minimum halo mass to host dusty galaxies is $\log M_{\rm min}/M_{\sun} \sim 10.8 \pm 0.6$.

\begin{figure*}
\begin{center}
        \includegraphics[width=19cm]{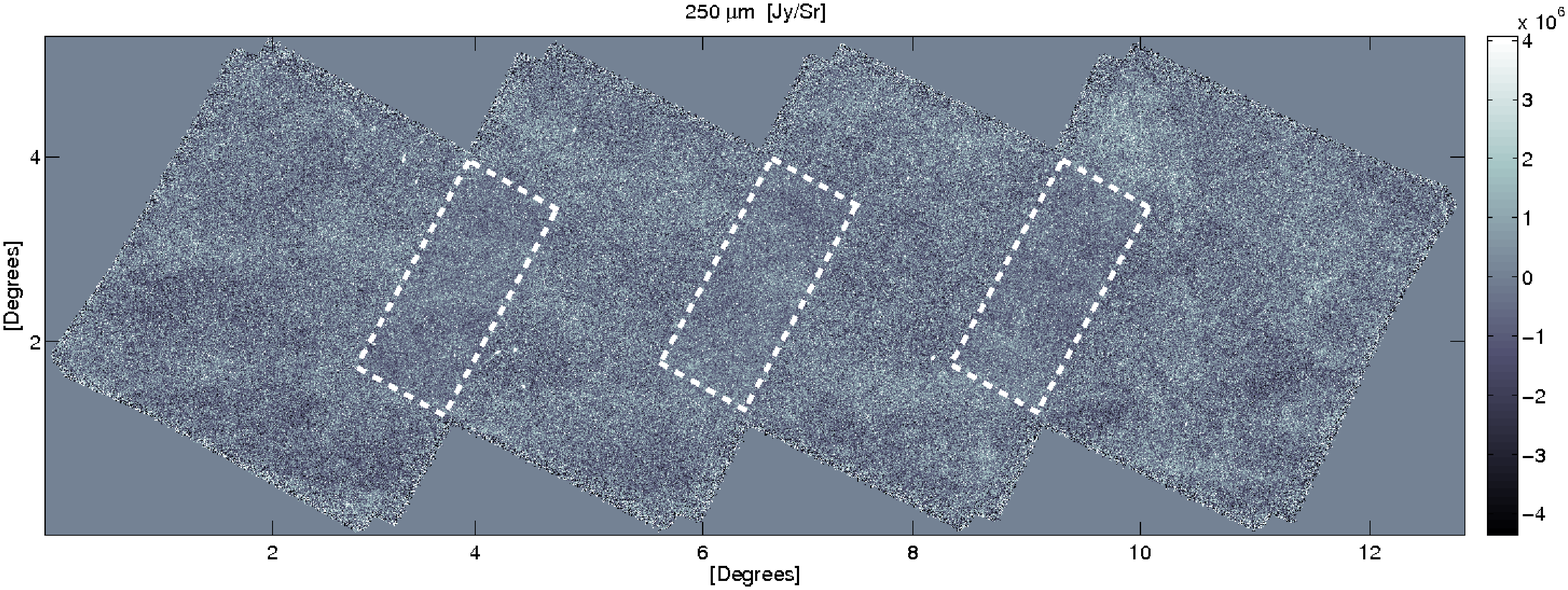}
        \includegraphics[width=19cm]{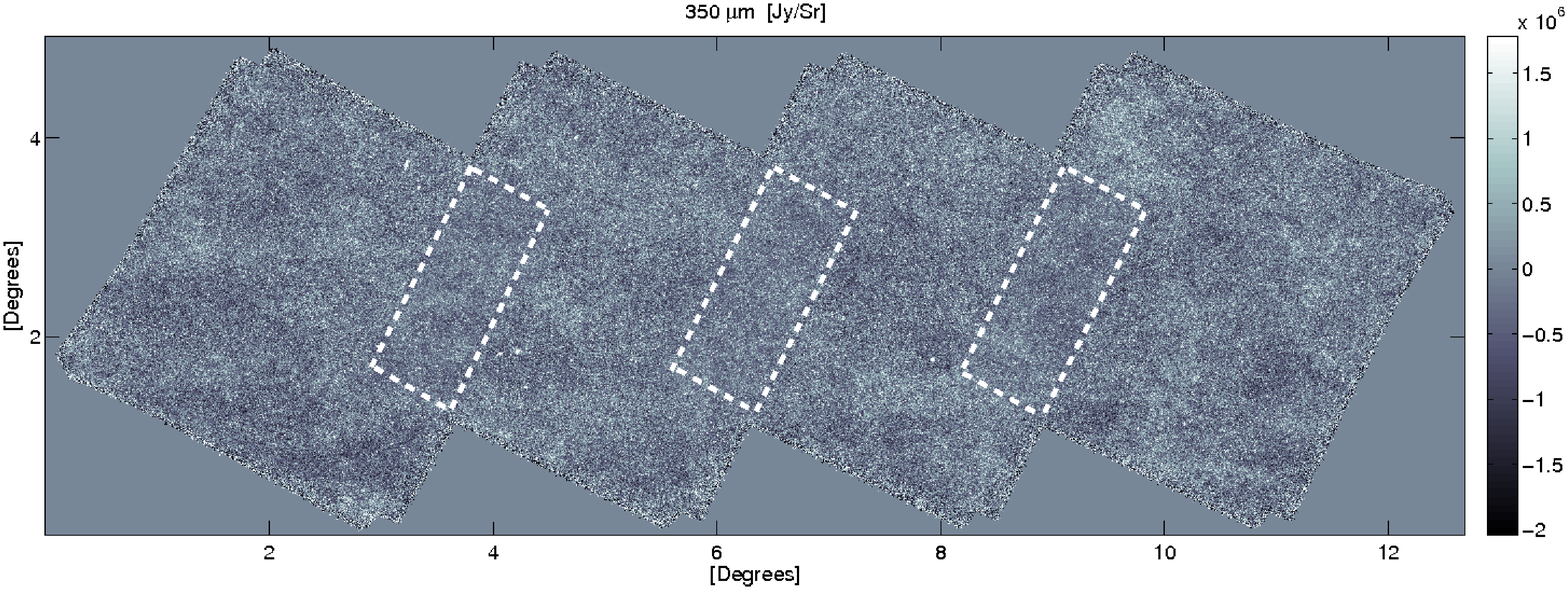}
        \includegraphics[width=19cm]{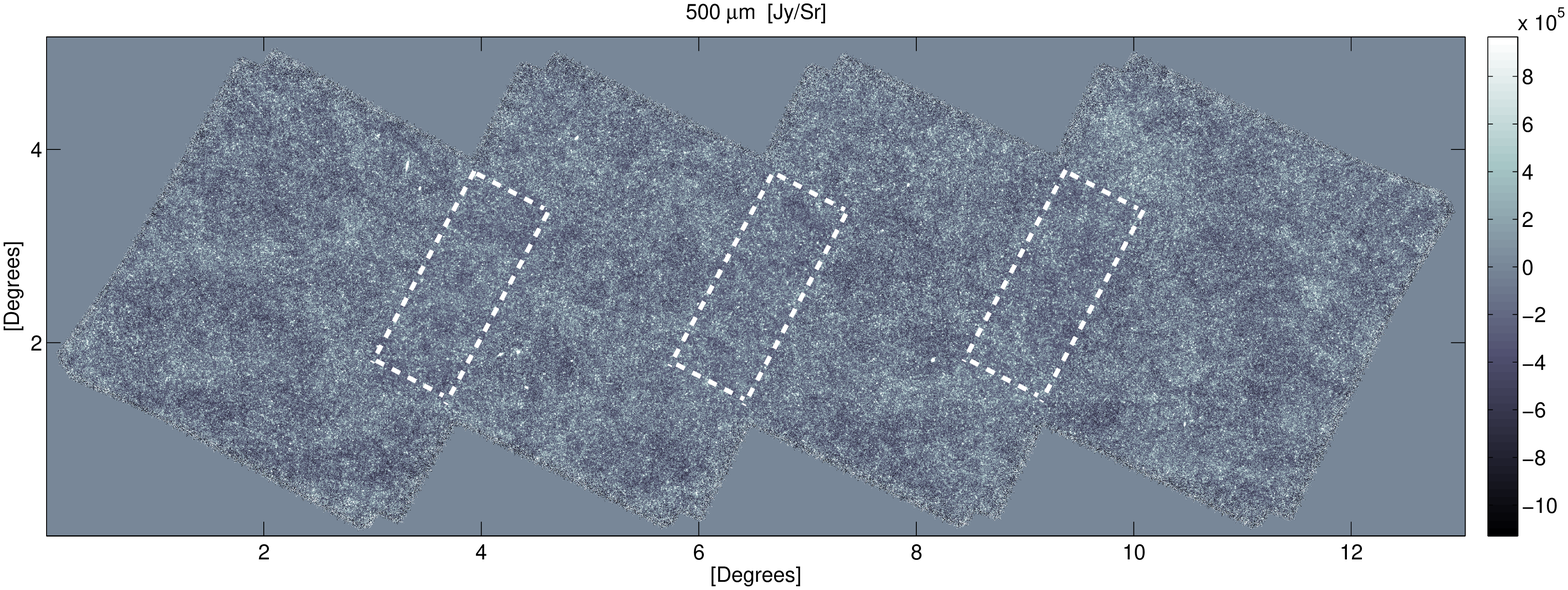}
\end{center}
    \caption[width=3in]{The {\it Herschel}-ATLAS GAMA-15 maps at 250 (top), 350 (middle) and 500 (bottom) $\mu$m with the three overlap
    regions used for the angular power spectrum measurements highlighted in dashed lines.}
    \label{fig:4tile_map}
\end{figure*}

The angular power spectrum of CFIRB, in principle, captures the spatial distribution of the background intensity, regardless of whether the emission is from
individual point sources or from smoothly varying diffuse sources, such as intracluster and intrahalo dust. Thus the angular power spectrum should be a sensitive probe
of the total dust content in the Universe. The existing estimes of the dust abundance from direct emission measurements make use of 
the sub-mm luminosity (e.g., Dunne, Eales \& Edmunds 2003) or dust mass (e.g., Dunne et al. 2011) functions, they
are generally based out of extrapolations of the measured bright galaxy counts. The anisotropy power spectrum should capture the integrated emission from faint sources, especially at the flux
density scale that dominate the confusion noise.  Separately, other estimates of the cosmic dust abundance rely on the extinction of optical light,
especially with measurements that combine magnification and extinction of quasars behind samples of foreground galaxies \citep{Menard2010,Menard:2012am}. 
It will be helpful to compare our direct emission measurement of the dust abundance with the extinction-based estimates since any differences
can allow us to understand the important of galaxies with hot dust that could be missed in SPIRE maps. 
We make use of a halo model to interpret the anisotropy power spectrum  with the goal of measuring $\Omega_{\rm dust}(z)$, the cosmic abundance of dust relative to the critical density, as a function of redshift.

To enable these measurements we make use of the wide field ($\sim$ 45 deg$^2$) maps of {\it
Herschel}-ATLAS (Eales et al. 2010)  in the three GAMA areas along the equator, and select a single area
that has the least Galactic cirrus confusion. This GAMA-15 field involves 4 independent blocks of
about 14 deg$^2$, each overlapping with the adjacent blocks by about 4 deg$^2$.
We make use of the three overlap
areas between the blocks to measure the power spectra at 250, 350 and 500\,$\mu$m. The
final power spectrum is the average of the individual power spectra of each of 
the overlapping regions. While this forces us to make a measurement over a smaller area than the
total survey area, our power spectrum measurement has the advantage that with two sets of cross-linked
scans we can make independent measurements of the noise power spectrum. 

Our measurement approach is similar to that used for HerMES power spectra
measurements (Amblard et al. 2011; Viero et al. 2012)  using multiple scans to generate jack-knives of data to test
the noise model. The total area used in HerMES measurements is about 12 and 60 deg$^2$, respectively, in Amblard et al. (2011)
and Viero et al. (2012). However, H-ATLAS covers about 120 deg$^2$ in all three GAMA fields
A measurement of the power spectrum in the whole of H-ATLAS GAMA areas requires an assumption about the noise power
spectrum, since in regions with only one orthogonal scan or a single cross-linkeds scan, we are not able to separate the noise
from the signal with data alone. In a future paper, we will present the power spectrum of the whole area 
using a noise model that is independently tested on various datasets to improve the confidence in separating noise
in single cross-link scans. For now, we make use of two cross-link scans for cross-correlations and auto-correlations to
separate noise and sky signal.

This paper is organized as follows. In Section~\ref{sec:mapmaking}, we briefly review how
250, 350 and 500\,$\micron$ maps for the GAMA-15 field were constructed using HIPE \citep{Ott:2010jz}
from raw time streams.  In Section~\ref{sec:powerspec}, we discuss how the auto
and cross-correlation functions for each of the three fields were estimated, corrected, and
assigned errors. The final power spectra are presented in Section~\ref{sec:powerspec_results}.
The halo model used to fit the data and the luminosity function is discussed in
Section~\ref{sec:halomodel}. Finally, in Section~\ref{sec:results} \& \ref{sec:conclusion} we present 
our results and their implications, discuss future follow up work and give our concluding thoughts.

\begin{figure*}
\begin{center}
        \includegraphics[scale=0.45]{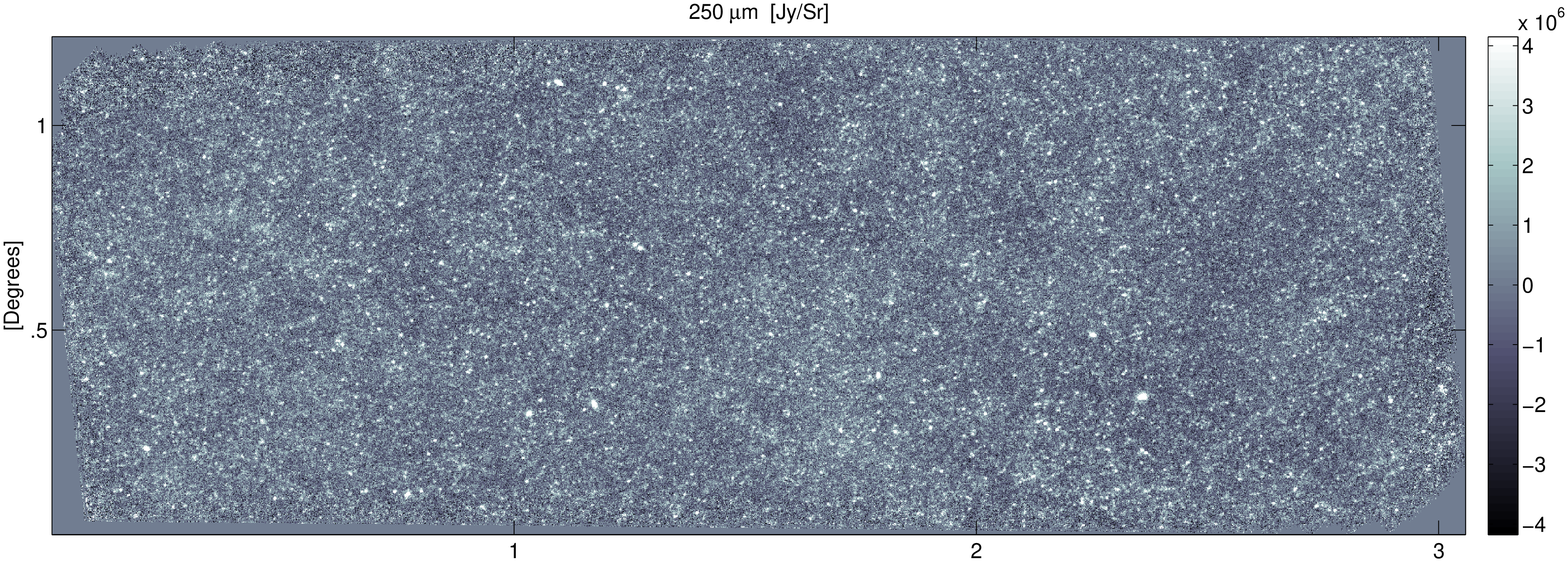}
        \includegraphics[scale=0.45]{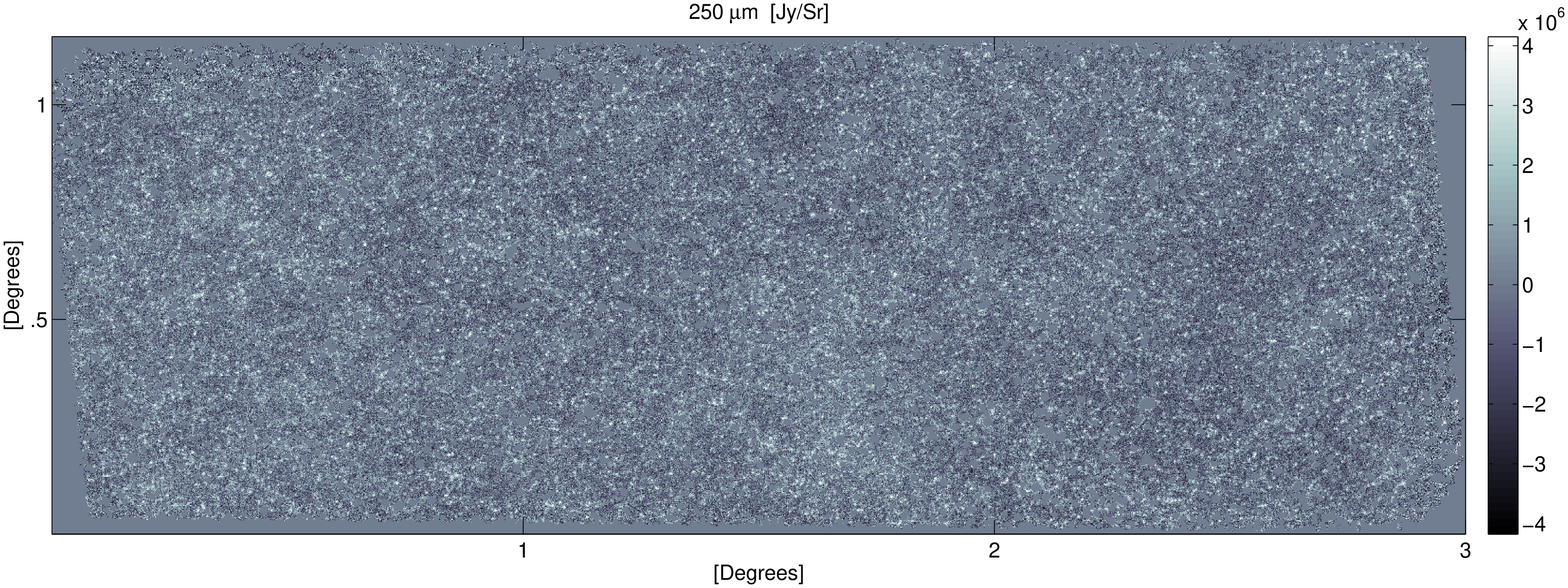}
\end{center}
    \caption[width=3in]{The left overlap region in Fig.~1 at 250\,$\mu$m {\it Herschel}-ATLAS showing details of the background 
intensity variations without (top) and with (bottom) $S>50$ mJy the bright source mask applied. This mask removes a substantial number of low-z bright galaxies
detected in the areas used for the fluctuation study.}
    \label{fig:overlap}
\end{figure*}

%\begin{figure}
%    %\begin{center}
%        \includegraphics[scale=0.5]{noise_all}
%    %\end{center}
%    \caption[width=3in]{The noise bias for the GAMA-15 $250 \mu m$
%field as calculated from the overlap regions. The blue xs show the mean
%auto-spectra from all 3 overlap regions. The red xs show the mean
%cross-correlations for these same overlap regions.  The black circles
%show the mean noise bias and error bars come from the standard deviation
%of the 3 overlap regions.  The cyan curve shows the extrapolated noise
%bias removed from scales larger than the overlap regions.}
%\label{fig:noise_bias}
%\end{figure}

\section{Map Making}
\label{sec:mapmaking}

For this work we generate SPIRE maps using the MADmap \citep{Cantalupo:2009if} algorithm that is available within HIPE.
The timeline data were reduced internally by the
H-ATLAS team using HIPE version 8.2.0 \citep{Pascale:2011br}. The timelines were calibrated with corrections applied for the
 temperature-drift and deglitched both manually and automatically.  Astrometry corrections were also
applied to the timelines using offsets between SDSS sources and the cross-identifications \citep{Smith:2011fi}.
In addition, a scan-by-scan baseline polynomial remover was applied to remove gain variations
leading to possible stripes.

The map-maker, MADmap, converts the timeline data $d(t)$
\begin{equation}
d(t) = n(t) + A(p,t) \times s(p) \, ,
\end{equation}
with noise $n(t)$ and sky signal  $s(p)$, given the pointing matrix $A(p,t)$ between pixel and time domain
to a map by solving the equation
\begin{equation}
m = (A^{\mathrm{T}}N^{-1}A)^{-1} A^\mathrm{T}N^{-1}d \, .
\end{equation}
Here $N$ is the time noise covariance matrix and $m$ is the pixel domain
maximum likelihood estimate of the noiseless signal map given $N$ and $d$. We refer the reader to
\citet{Cantalupo:2009if} for more details of MADmap.

The final maps we use for this work consist of four partially overlapping tiles, each containing two sets
of 96 scans in orthogonal directions (Fig.~1). 
The pixel-scale for the 250, 350 and 500\,$\mu$m maps is 6\arcsec, 8.333\arcsec\, and 12\arcsec, respectively, corresponding to $1/3$ of the beam size.

In regions where the tiles do not overlap, the map at each wavelength
consists of a single scan each in the two orthogonal scan directions. In the overlap region, we have two scans in each direction. As discussed below,
we are able to estimate the noise and signal power spectra independent of each other using the auto-correlations of the combined 4-scan map and the cross-correlations involving
various jack-knife combinations. In Fig.~2 we show an example overlap region.

\begin{figure}
    \begin{center}
    \includegraphics[scale=0.4]{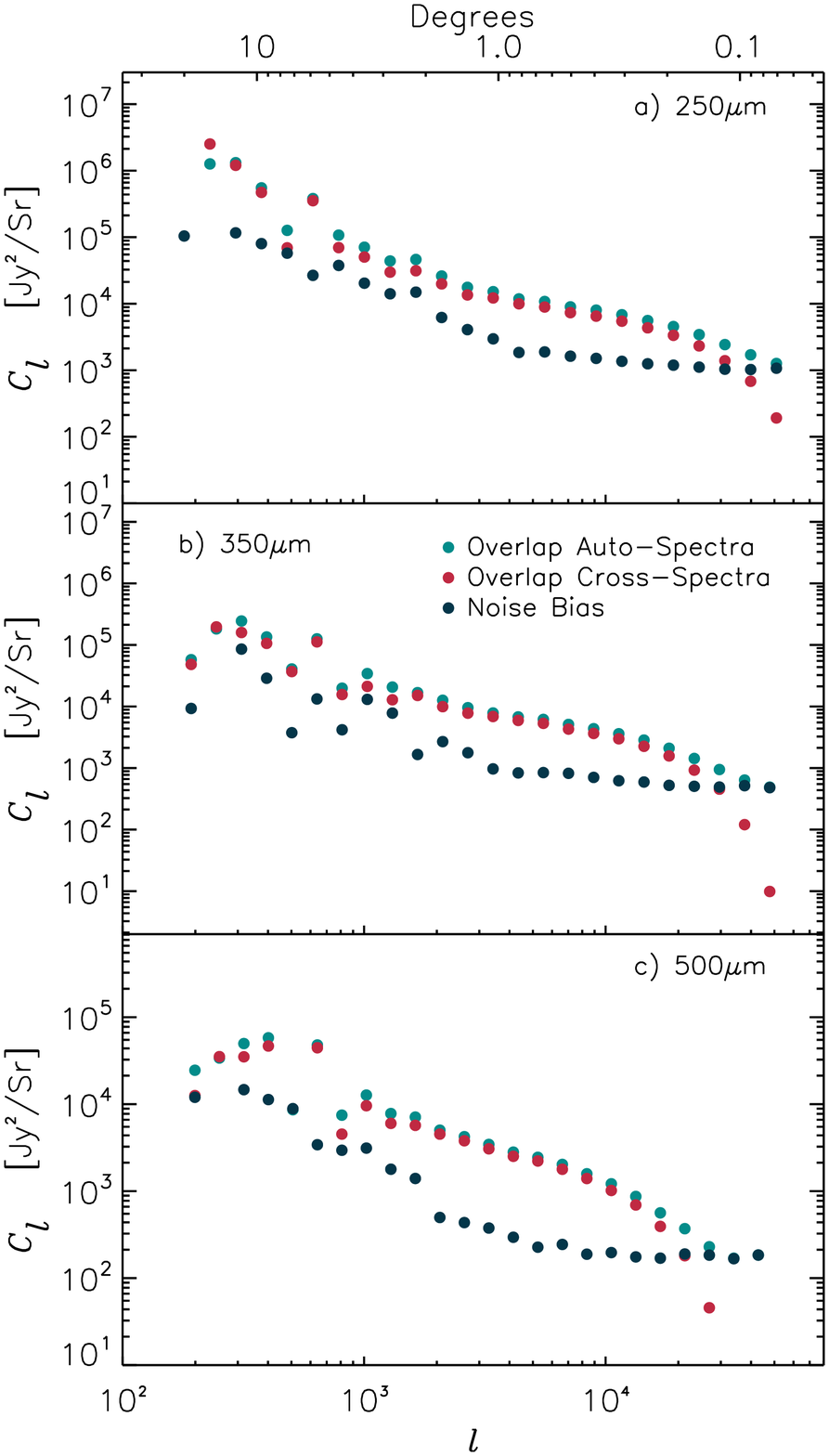}
    \caption[width=3in]{The raw $C_l$ of the GAMA-15 overlap regions at 250 (top), 350 (middle) and 500 (bottom) $\mu$m, respectively. The
green points show the auto-power spectra computed from overlap regions using all 4 scans. This power spectrum is a combination of the real sky anisotropy
power spectrum and the instrumental noise. We estimate the sky signal independent of noise by creating two sets of maps for each of the three overlap regions with two
orthogonal scans each and then taking the cross power spectrum (red points) of those independent maps.
The difference of these two spectra shows the instrumental noise power spectrum (black points).}
\label{fig:noise_all}
    \end{center}
\end{figure}

\section{Power Spectra}
\label{sec:powerspec}

We now discuss the measurement of angular anisotropy power spectra in each of the three SPIRE bands.
To be consistent with previous measurements of the SPIRE angular power spectrum
\citep{Amblard:2011gc} our maps are masked by taking a 50 mJy/beam flux cut and then convolving with
the point response function. Such a flux cut, through a mask that removes the bright galaxies, also
minimizes the bias coming from those bright sources by reducing shot-noise effects.
The same mask also includes a small number of pixels that do not contain any useful data, either due to scan strategy or data corruption.
The combined mask removes roughly 13, 12 and 15\% of the pixels at 250, 350 and 500\,$\mu$m,
respectively. The fractions of masked pixels are substantially higher than the fractions of Amblard et al. (2011)
of 1 to 2\% as the ATLAS GAMA-15 field has a large density of $z < 0.1$ spiral galaxies over its
area relative to more typical extragalactic fields used in the Amblard et al. (2011) study. These galaxies tend to be brighter,
especially at 250\,$\mu$m. While the fraction masked is larger, the total number of pixels used for this study is comparable to Amblard et al. (2011) with $2.9\times10^6$, $1.5 \times 10^6$ and $7.0\times10^5$ at 250, 350
and 500\,$\micron$ in each of the three overlap regions.

To measure the power spectrum in the final set of maps, we make use of 2D Fourier transforms.  In general this is done with
masked maps of the overlap regions, denoted $M_1$ and $M_2$ in real space. If we denote the 2D Fourier
transform of each map as $\widetilde{M}_1$
and $\widetilde{M}_2$, the power spectrum, $C_l$, formed for a specific $l$ bin between between $l$-modes $l_1$ and $l_2$, is the mean of the squared
Fourier modes $\tilde{M}_1\tilde{M}^*_2$ between $l_1$ and $l_2$.  The same can be used to describe the auto power spectra, but with
$M_1 = M_2$.

The raw power spectra are summarized in Fig.~3.
Here, we show the auto spectra in the total map, as well as the cross spectrum with maps made with half of the time-ordered data in each map. The difference of the two
provides us with an estimate of the instrumental noise.
At small angular scales (large $\ell$ values) the noise follows a white-noise power spectrum, with
$C_l$ equal to a constant. At large angular scales, the detectors show the expected $1/f$-type of
noise behavior, with the noise power spectrum rising as $C_l\propto l^{-2}$.
We fit a model of the form
\begin{equation}
N_l = N_0 \left[\left(\frac{l_0}{l}\right)^2 +1 \right] \, ,
\end{equation}
and determine the knee-scale of the $1/f$ noise and the amplitude of noise power spectra. The  noise
values are $N_0=1.2 \times 10^3, 5.3 \times10^2$ and $1.8\times10^2$ Jy$^2$/sr at 250, 350 and 500\,$\mu$m, comparable to the
detector noise in the 4-scan maps of the Lockman-hole used in Amblard et al. (2011). The knee at which $1/f$ noise becomes important is $l_0=3730,2920$ and 3370, comparable to the expected knee at a wavenumber of 0.15 arcmin$^{-1}$
given the scan rate and the known properties of the detectors \citep{Griffin:2010hp}.

The raw spectra we have computed directly from the masked maps are contaminated by several different effects
that must be corrected for. These issues are: the resolution damping from the instrumental beam, the
filtering in the map-making process, and
the fictitious correlations introduced by the bright source and corrupt pixel mask. Including these effects, we can write the measured
power spectrum as
\begin{equation}
C_l' = B^2(l) T(l) M_{ll'} C_{l'} \, ,
\end{equation}
where $C_l'$ is the observed power spectrum from the masked map, $B(l)$ is the beam function measured in a map, $T(l)$ is the map-making transfer function, and
$M_{ll'}$ is the mode coupling matrix resulting from the mask. Here, $C_{l'}$ is the true sky power spectrum and is determined by inverting the above equation.

We now briefly discuss the ways in which we either determine or correct for the effects just outlined.

\begin{figure}
    \begin{center}
    \includegraphics[scale=0.48,clip]{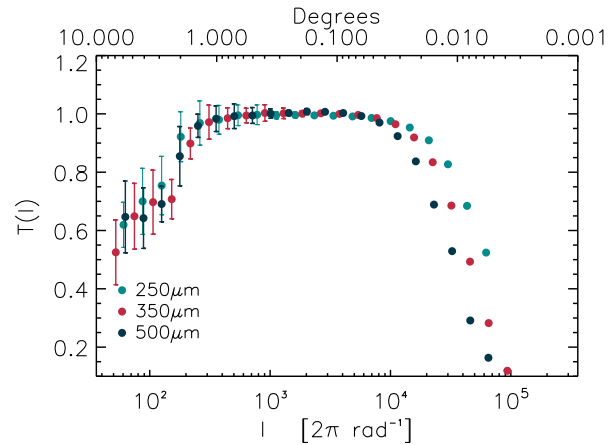}
    \caption[width=3in]{Map-making transfer function $T(l)$ for the MADmap map making tool used for
    the GAMA-15 field anisotropy power spectrum measurement. The uncertainties in the
transfer function are calculated from 100 random realizations of the sky as described in Section~3.1.}
\label{fig:transfer}
    \end{center}
\end{figure}

\begin{figure}
    \begin{center}
    \includegraphics[scale=0.48,clip]{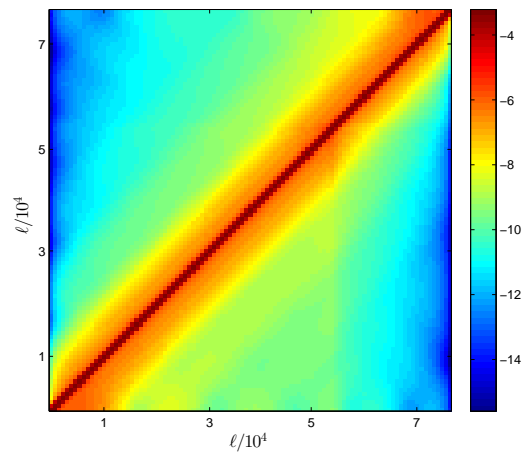}
    \caption[width=3in]{An example inverse-mode coupling matrix $M_{ll'}^{-1}$ for one of the overlap regions (log scale). }
\label{fig:mode}
    \end{center}
\end{figure}

\subsection{The Map-Making Transfer Function}

Due to the finite number of detectors, the scan pattern, and the resulting analysis technique to
convert timeline data into a map,
the map we produce is not an exact representation of the sky. The modifications are associated with
the map-making process, relative to the true sky,
are described by the transfer function $T(l)$. We determine this by making 100 random realizations of the sky using Gaussian random fields derived from  a
first estimate of the H-ATLAS power spectrum. We sample those skies using the same timeline data as the actual observations and analyze the simulated timelines
with the same data reduction and map-making HIPE scripts for the actual data. We then compute the average of the ratio between the estimated power spectra
and the input spectrum. This function is then the transfer function associated with polynomial
filtering and the map-making process.
This transfer function, like the beam, represents a multiplicative correction to the data.  We divide the estimated power spectrum of the data by this transfer function to
remove the map-making pipeline processing effects.

In Fig.~4 we show the transfer functions at 250, 350 and 500\,$\mu$m with 68\% error bars taken from the standard
deviation of 100 simulations. The transfer function is such that it turns over from 1 at both large angular scales, corresponding to roughly the scale of an individual
scan  length, and the beam scale. The large-scale deviation,  which is wavelength independent, is due to the polynomial removal
 from each timeline of data, while the turnover at small angular scales is due to the cut-off
imposed by the instrumental point response function or the beam. The transfer function is more uncertain at the large angular scales due to
the finite number of simulations and the associated cosmic variance resulting from the field size.
Given its multiplicative nature, errors from this transfer function are added in quadrature with rest of the errors.

\subsection{The Beam}
                              
Following Amblard et al. (2011), the beam function is derived from Neptune observations of SPIRE. 
The Neptune timeline data are analyzed with the same pipeline and our default map maker in HIPE. The resulting beam functions are similar to those of Amblard et al. (2011)
and we find no detectable changes resulting from the two different map makers between this work and the SMAP \citep{Levenson2010} pipeline of the SPIRE Instrument Team used in Amblard et al. (2011) and Viero et al. (2012).
This is primarily due to the fact that the beam measurements involve a large number of scans and Neptune is several orders of magnitude brighter than the extragalactic confusion noise. We interpolate the beam function
measured from Neptune maps in the same $\ell$ modes at which we compute our anisotropy power spectra. 
This beam transfer function, $B(l)$, at each of the wavelengths represents a multiplicative correction to the data.
Similar to Amblard et al. (2011), we compute the uncertainty in the beam function by computing the standard deviation of
several different estimates of the beam function by subdividing the scan data to 4 different sets. 
The error on the beam function in Fourier space is propagated to the final error and is added in quadrature with rest of the errors.

\subsection{Mode-Coupling Matrix}

The third correction we must make to the raw power spectrum involves the removing of 
fictitious correlations between modes introduced by the bright sources and contaminated or zero-data pixel mask.  
Due to this mask the 2D Fourier transforms are measured in maps with holes in them. In the power spectrum these holes
result in a Fourier mode coupling that biases the power spectrum lower at large angular scales and higher at smaller angular scales.
This can be understood since the modes at the largest angular scales, like the mean of the map, are
broken up into smaller scale modes with any non-trivial mask.

To correct for the mask we make use of the method used in \citet{Cooray2012}.
The method involves capturing 
the effects of the mask on the power spectrum into a mode-coupling matrix $M_{ll'}$. The inverse of the mode coupling matrix then
removes the contamination and corrects the raw power spectrum to a power spectrum that should be measurable in an unmasked sky.
The correction both restores the power back to the large 
angular scale modes by shifting the power away from the small angular scale modes, especially those at the modulation scale
introduced by the mask. 

To generate $M_{ll'}$ we apply the mask to a map consisting of a Gaussian realization of a single $l$-mode 
and take the power spectrum of the resulting map.  This power spectrum represents the shuffling of power the mask performs on this specific $l$-mode 
among the other $l$-modes.  This process is repeated for all $l$-modes and these effects of the mask on each mode are then stored in a matrix.  
This matrix, $M_{ll'}$ now represents the transformation from an unmasked to a masked sky by construction.  By inverting this matrix, shown in Fig. 5, 
we are left with the transformation from a masked to an unmasked sky removing the fictitious couplings induced by the mask applied to the raw power spectra.  
The matrix $M_{ll'}$ behaves such that  in the limit of no $l-$mode coupling $M_{ll'} =f_{\rm sky}
\delta_{ll'}$ where  $f_{\rm
sky}$ is the fraction of the sky covered. Thus in the limit of partial sky coverage the correction  becomes the standard formula with
$C_l'=f_{\rm sky} C_l$.   For more details, including figures demonstrating the robustness of the
method, we refer the reader to \citet{Cooray2012}.

%\begin{table}
%\begin{center}
%\begin{tabular}{ |c | c| c|  }
%\hline
%  Wavelength & Atlas (Jy$^2$/Sr) \\
%\hline
%  250 & $6382 \pm 447$ \\
%  350 & $4381 \pm 339$ \\
%  500 & $1893 \pm 158$ \\
%\hline
%\end{tabular}
%\label{tab:Poisson}
%\caption[width=3in]{Best fit Poisson Noise level for each field from the halo model. The errors represent the 68\% confidence region.}
%\end{center}
%\end{table}

\begin{figure*}
    \begin{center}
    \includegraphics[scale=0.47,clip]{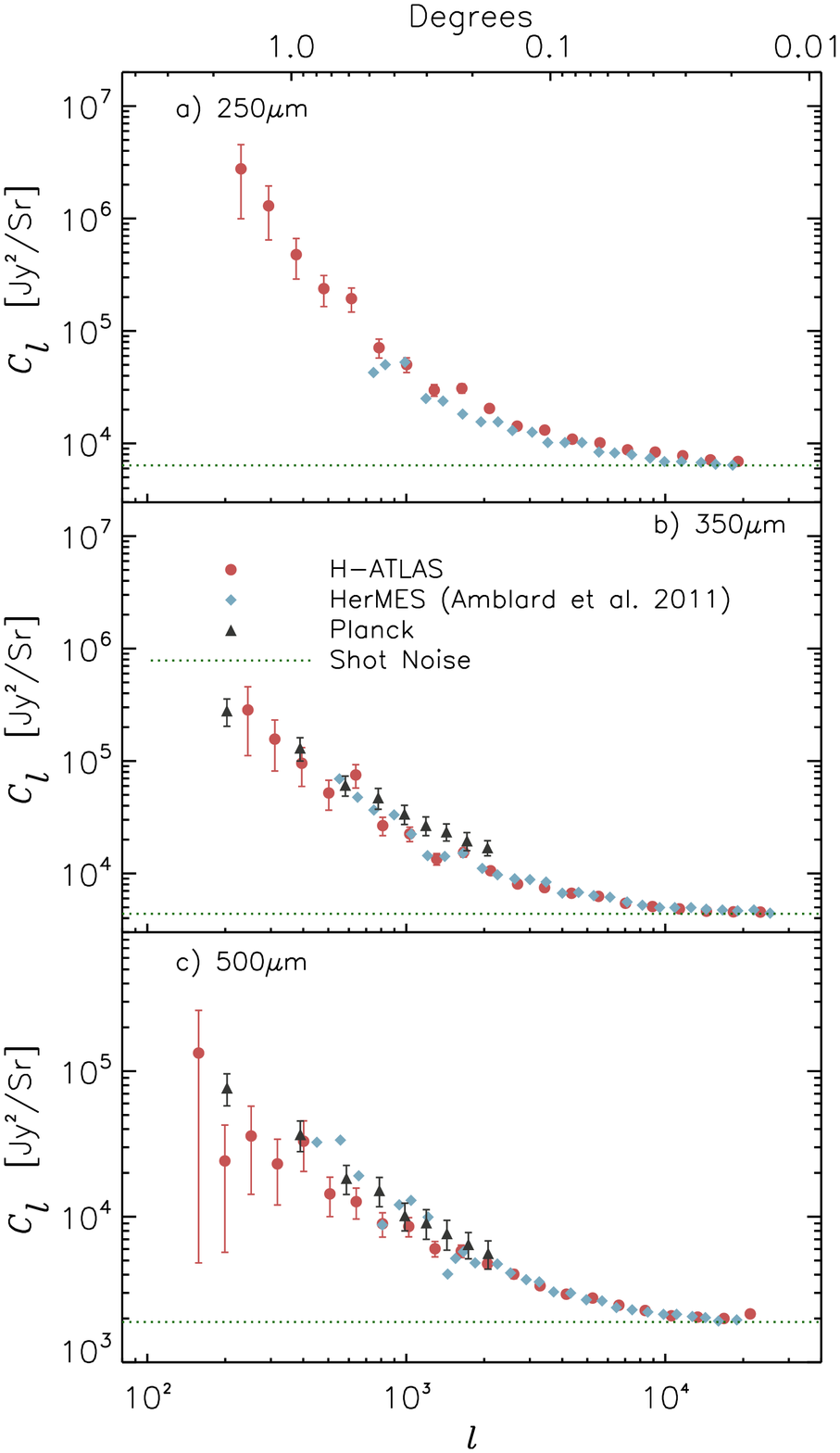}
    \includegraphics[scale=0.47,clip]{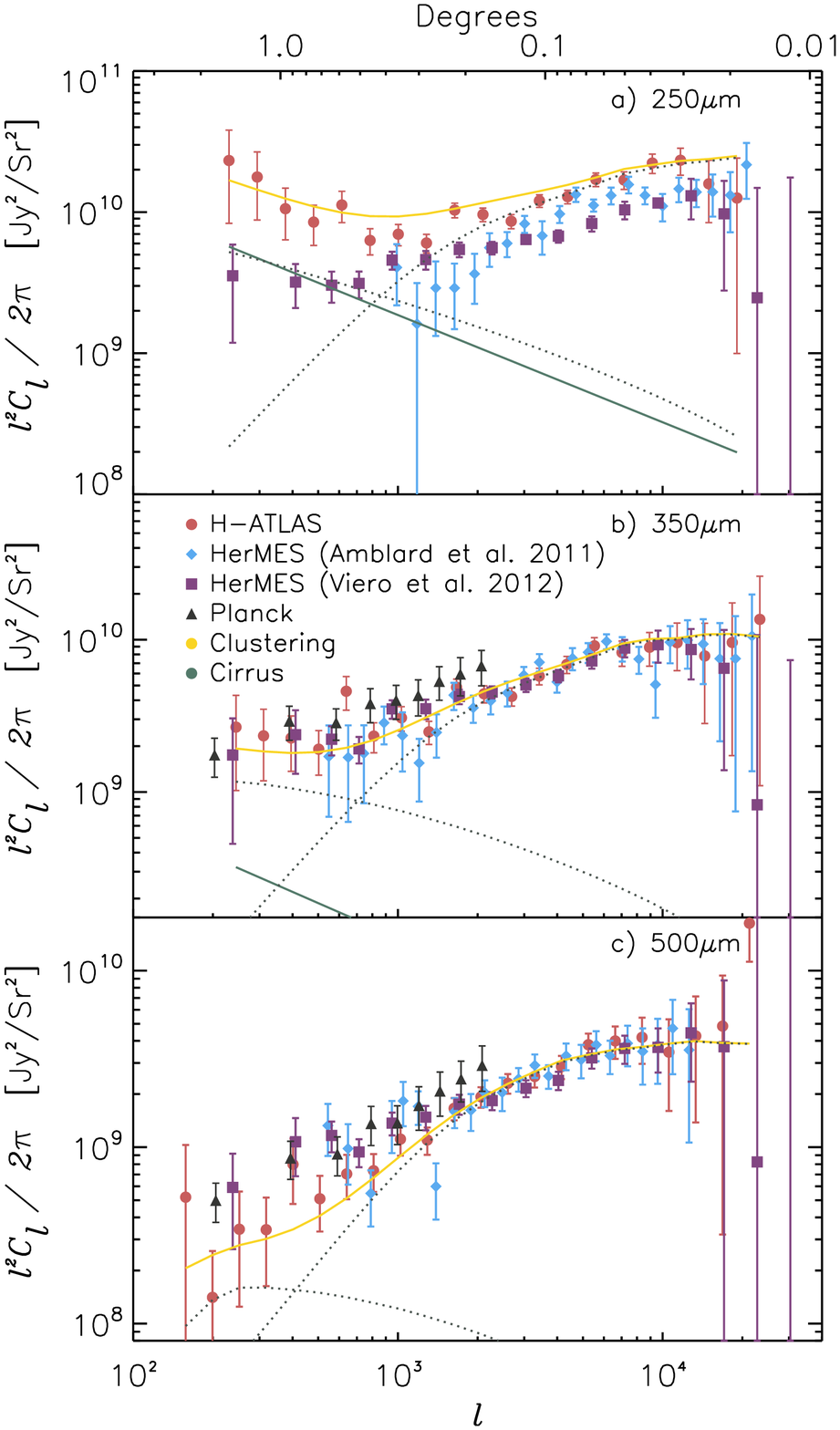}
    \end{center}
    \caption{Final angular power spectra  of CFIRB anisotropies in the H-ATLAS GAMA-15 field at 250 (top), 350 (middle) and 500 (bottom) $\mu$m. In the left panels,
the power spectra are plotted as $C_l$ prior to the removal of the shot-noise term. Here we compare
the power spectra measured with H-ATLAS data to Planck and previous {\it Herschel} results from
HerMES. In the right panels we show the power spectra as $l^2C_l/2\pi$ after removing the shot-noise
level at each of the wave bands.
We add the uncertainty associated with the shot-noise level back to the total error budget in quadrature. This results in the increase in errors at high multipoles
or small angular scales. The curves show the best-fit model separated into 1 and 2-halo terms (see
text for details) and the total (orange line). The solid line
that scales roughly as $l^2C_l \sim l^{-0.9}$ is the best-fit Galactic cirrus fluctuation power
spectrum. Due to the high cirrus fluctuation amplitude and clustering, the H-ATLAS power spectrum in
the GAMA-15 field at 250\,$\mu$m is higher than the existing HerMES results, while the measurements
are generally consistent at 350 and 500 $\mu$m.}
\label{fig:powerspec}
\end{figure*}

\section{Power Spectrum Results}
\label{sec:powerspec_results} 

The final power spectrum $C_l$ at each of the three wavelengths is shown in Fig.~6 (left panels). The final error bars account for the uncertainties associated with the (a)
beam, (b) map making transfer function, (c) instrumental or detector noise (Fig.~3), and the cosmic variance associated with the finite sky coverage of the field.
In Fig.~6 we  compare these final H-ATLAS GAMA-15 power spectra with measurements of the CFIRB anisotropy power spectrum measurements
from HerMES (Amblard et al. 2011) and Planck (Planck collaboration 2011) team measurements. We find general agreement, but we also
find some differences. At 250\,$\mu$m, we find the amplitude to be larger than the existing SPIRE
measurements of the power spectrum at 250\,$\mu$m in HerMES
while the measurements are more consistent at 350 and 500\,$\mu$m. 
We attribute this increase to the wide coverage of H-ATLAS and the presence of a large surface density of galaxies at low redshifts. While most of these galaxies are masked
we find that the fainter population likely remains unmasked and contributes to the increase in the power that we have seen. This conclusion is also consistent with the
strong cross-correlation between detected SPIRE sources in GAMA fields of H-ATLAS and the SDSS redshift survey (e.g., Guo et al. 2011).
The difference between {\it Herschel}-SPIRE measurements and Planck measurements are discussed in Planck collaboration (2011)
and we refer the reader to that discussion. We continue to find differences between our measurements and Planck power spectra at 
350\,$\mu$m, even with Planck data corrected for the frequency differences and other corrections associated with the source mask, 
as discussed in Planck collaboration (2011).

Note that the power spectra in the left panels of Fig.~6 asymptote to a $C_l \sim$ constant. This is the shot-noise coming from the Poisson behavior of the sources.
In Fig.~6 (right panels) we show the final power spectra plotted
as $l^2C_l/2\pi$, with the Poisson noise removed at each band. They now reveal the underlying clustering of
submillimeter galaxies. 
With sources masked down to 50 mJy, our shot-noise amplitudes are 
$6700 \pm 140$, $4400 \pm 130$ and $1900 \pm 90$ Jy$^2$/sr at 250, 350 and 500\,$\mu$m, respectively (see Table~1).
We determine the Poisson noise uncertainties based on the overall fit to $C_l$ measurements at the three highest $\ell$-bins. 

For comparison to our shot-noise values, the shot-noise values of Amblard et al. (2011) are $6100
\pm 120$, $4600 \pm 70$ and $1800 \pm 80$ Jy$^2$/sr at 250, 350 and 500\,$\mu$m, respectively.
While the shot-noise values are consistent at 350 and 500\,$\mu$m, we find an increased shot-noise
amplitude at 250\,$\mu$m, consistent with the higher amplitude of the clustering part of the power spectrum.
In addition to Planck and Amblard et al. (2011) HerMES
measurements, in Fig.~6 (right panels) we also compare our measurements to more recent Viero et al. (2012)
HerMES  measurements. At 350 and 500\,$\mu$m the difference between all of Herschel-SPIRE
measurements and Planck is clear. 

At 250\,$\mu$m, we find that our measurements have a higher amplitude at all angular scales
relative to previous  SPIRE measurements. At large angular scales, we find that the increase is coming from the higher intensity of cirrus in our GAMA-15 fields (Bracco et al. 2011). The cirrus properties as measured from the
power spectra are discussed in Section~6.1. As part of the discussion related to our results on the galaxy distribution that is contributing the far-IR background power spectrum 
(Section~6.2), we will explain the difference between the HerMES and H-ATLAS power spectrum at 250 $\mu$m as due to an excess of low-redshift galaxies in the H-ATLAS GAMA-15 field (Rigby et al. in prep). The measurements shown in Fig.~6 right panels constitute our final CFIRB power spectrum measurements in the H-ATLAS GAMA-15 field. These power spectra values are tabulated in Table~2. 
We now discuss the model used for the interpretation leading to the best-fit model lines shown in Fig.~6 (right panels).

\section{Halo Modeling of the CFIRB Power Spectrum}
 \label{sec:halomodel}

To analyze the H-ATLAS GAMA-15 power spectra measurements we implement the conditional luminosity
function (CLF) approach of \citet{Giavalisco:2000xs,
Lee:2009,DeBernardis:2012bf}. We recall below the main features of the
model and refer the reader to these works for more details. The goal is to work out the relation between IR luminosity and halo masses of the galaxies
that are contributing to the CFIRB power spectrum. We populate halos with the best-fit $L_{\rm IR}(M)$ relation from the data and use that to determine the
abundance of dust ($\Omega_{\rm dust}$) in the Universe. The CLF approach proposed here improves over several assumptions that were made in Amblard et al. (2011) to interpret the first Herschel-SPIRE anisotropy
power spectrum measurements.

First, the probability density for a halo or a sub-halo of mass $M$ to host a galaxy with IR luminosity $L$ is modeled as a normal distribution with:
\begin{eqnarray}
P(L|M)=\frac{1}{\sqrt{2\pi}\sigma_L(M)}\exp\left[-\frac{(L-\bar{L}(M))^2}{2\sigma_L(M)^2}\right] \, .
\label{pml}
\end{eqnarray}
The relation between the halo mass and the average luminosity $\bar{L}(M)$ is expected to be an increasing function of the mass with a characteristic mass scale $M_{0l}$ and we can write (see \citet{Lee:2009})
\begin{eqnarray}\label{lm_relation}
\bar{L}(M)=L_{0}\left(\frac{M}{M_{0l}}\right)^{\alpha_l}\exp\left[-\left(\frac{M}{M_{0l}}\right)^{-\beta_l}\right] \, .
\end{eqnarray}
As already discussed by  \citet{Lee:2009} these parameterizations do not have a specific physical motivation,
except for the requirement that the luminosity increases as an increasing function of the halo mass and offer the advantage that one can
 explore a large range of possible shapes for the luminosity-mass relation. While there is no motivation  to use this specific form over another,  certain
models of galaxy formation do predict a $L(M,z)$ relation and our results based out of the model-fits to CFIRB power spectrum can be compared to those model predictions.
In particular, the model of Lapi et al. (2011) predicts $L(M,z) \propto  M(1+z)^{2.1}$, while the cold-flow accretion model of Dekel et al. (2009)
predicts $L(M,z) \propto  M^{1.15}(1+z)^{2.25}$.

The total halo mass function is given by the number density  of halos or sub-halos of mass $M$.
The contribution of halos $n_h(M)$  is taken to be the Sheth \& Tormen relation \citep{ShethTormen1999}.
The sub-halos term can be modeled through the number of sub-halos of mass $m$ inside a parent halo of mass $M_p$, $N(m|M_p)$. 
The total mass function is then written as
\begin{eqnarray}
n_T(M)=n_h(M)+n_{sh}(M) \, ,
\end{eqnarray}
where $n_{sh}(M)$ is the sub-halo mass function
\begin{eqnarray}
n_{sh}(M)=\int N(M|M_p)n_h(M_p)dM_p \, .
\end{eqnarray}
Here we parameterize $N(m|M)$ following the semi-analytical model of \citet{vandenBosch:2004zs}.

Neither the normalization nor the slope of the sub-halo mass function
are universal and both depend on the ratio between the parent halo mass and the non-linear mass scale,
$M_*$. $M_*$ is defined as the mass scale where the rms of the density field $\sigma(M,z)$ is equal to the
critical over-density required for spherical collapse $\delta_c(z)$. The contribution of central
galaxies to the halo occupation distribution (HOD) is simply the integral of $P(L|M)$ over all 
luminosities above a certain threshold $L_0$, either fixed by the survey or a priori selected so that
\begin{eqnarray}
\langle N_c(M)\rangle_{L\ge L_{min}}=\int_{L_{min}}P(L|M)dL \, ,
\end{eqnarray}
which, in the absence of scatter, reduces to a step function $\Theta(M-M_0)$, as expected. 
Note that all integrals over the luminosity $L$ also have a redshift-dependent 
cut-off at the upper limit corresponding to the flux cut of 50 mJy that we used for the power spectrum measurement.
For the satellite galaxies,  the HOD is related to the sub-halos
\begin{eqnarray}
\langle N_{s}(M)\rangle_{L\ge L_{min}}=\int_{L_{min}}dL\int dmN(m|M) P(L|m)\,. \nonumber \\
\end{eqnarray}
The total HOD is then
\begin{eqnarray}
\langle N_{\rm tot}(M)\rangle_{L\ge L_{min}}=\langle N_{h}(M)\rangle_{L\ge L_{min}}+\langle N_{sh}(M)\rangle_{L\ge L_{min}} \, . \nonumber \\
\end{eqnarray}

\begin{table}
\caption[width=3in]{Parameter values from MCMC fits to the H-ATLAS GAMA-15 angular power spectra
at $250$, $350$ and $500$\,$\mu$m.} 
\begin{center}
\begin{tabular}{llr}
\hline\hline
HOD & $\alpha_l$   								&	$0.69\pm0.04$\\
& $\beta_l$     											&	$0.09\pm0.05$\\
& $\log (L_{0}/L_{\odot})$							   &	$9.52\pm0.08$\\
& $\log (M_{0}/M_{\odot})$							   &	$11.5 \pm 1.7$\\
& $P_M$    						                &	$-2.9\pm0.4$\\
\hline
CFIRB SED & $T_{\rm dust}$ 							& $37\pm2$ K\\
& $\beta_{\rm dust}$							& unconstrained\\
\hline
Cirrus & $C_{250}^{l=230}$							& 	$3.5\pm1.3 \times 10^5$ Jy$^2$/sr		\\
& $C_{350}^{l=230}$							& 	$1.2\pm1.0 \times 10^4$ Jy$^2$/sr		\\
& $C_{500}^{l=230}$							& 	$1.1\pm0.9 \times 10^3$ Jy$^2$/sr		\\
& $T_{\rm cirrus}$                                                    &    $21.1 \pm 1.9$ K\\
& $\beta_{\rm cirrus}$                                               &  $2.9 \pm 0.8$ \\
\hline
Poisson & $SN_{\rm 250}$ 						&	$6700\pm140$ Jy$^2$/sr				\\
& $SN_{\rm 350}$ 						&	$4400\pm130$	  Jy$^2$/sr			\\
& $SN_{\rm 500}$ 						&	$1900\pm90$    Jy$^2$/sr				\\	
\hline
\end{tabular}

\end{center}
\label{tab:tab_fit}
\end{table}

\begin{figure}
    \begin{center}
    \includegraphics[scale=0.5,clip]{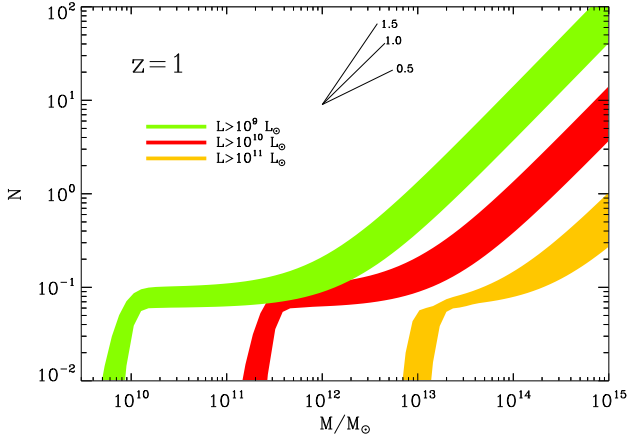}
   \caption[width=3in]{Best-fit halo occupation distribution and the $1\sigma$
    range at $z=1$ for three cases involving $L_{\rm IR} > 10^9, 10^{10}$ and 10$^{11}$ L$_{\sun}$.
 The three lines to the top show the different power-laws for comparison with the shape of the HOD. 
The satellite galaxies
    contribution has a slope $\sim1$ when $L_{\rm IR} \sim 10^9$ L$_{\sun}$.}
    \end{center}
    \label{fig:hod_range}
\end{figure}

\begin{figure}
    \begin{center}
    \includegraphics[scale=0.5,clip]{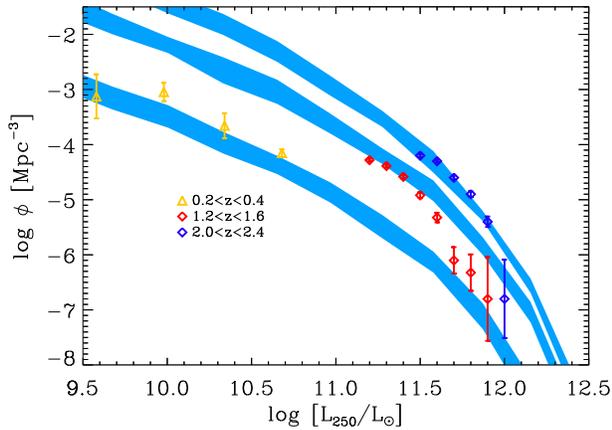}
   \caption[width=3in]{ Luminosity functions predicted by our model compared to data from
   \citet{Eales:2010vw} $(0.2< z< 0.4)$ and \citet{Lapi:2011ca} $(1.2< z< 1.6, 2< z< 2.4)$. The shaded region
   corresponds to the 68\% confidence level.}
    \end{center}
    \label{fig:LFs_new}
\end{figure}

\begin{figure}
    \begin{center}
        \includegraphics[scale=0.5,clip]{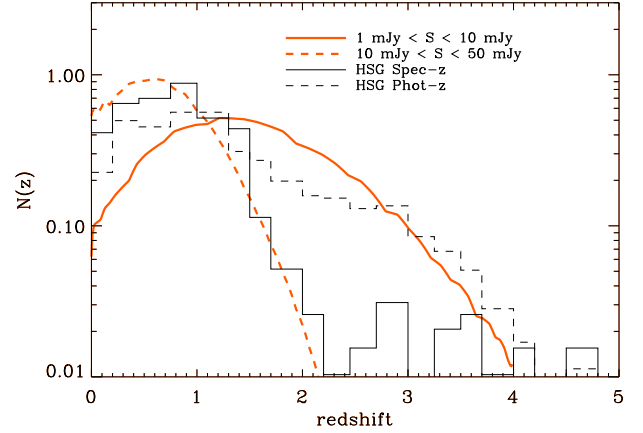}
            \caption{ Normalized redshift distributions of FIR-bright galaxies predicted by our
model for two different flux density cuts at 250 $\mu$ (thick solid line for $1\, {\rm mJy} < S < 10 \, {\rm mJy}$ and thick dashed line
for $10\, {\rm mJy} < S < 50 \, {\rm mJy}$). For comparison in corresponding thin lines we show the measured redshift
distributions for the Herschel-selected galaxies (HSGs) at the same flux density bins with optical spectra in \citet{Casey2012}.}
    \end{center}
    \label{fig:nz}
\end{figure}

We account for the possible redshift evolution of the luminosity-halo mass relation by introducing the parameter $p_M$
and rewriting the mass scale $M_{0l}$ as:
\begin{eqnarray}\label{mol_evolution}
M_{0l}(z)=M_{0l,z=0}(1+z)^{p_M} \, .
\end{eqnarray}
Under the assumption that the central galaxy is at the center of the halo and that the halo radial profile of satellite galaxies within dark matter halos follow that of the dark matter, given by the Navarro, Frenk and White (NFW) profile \citep{NFW}, we can write the 1-halo and 2-halo terms of the three-dimensional power spectrum:
\begin{eqnarray}\label{1h}
P^{1h}(k)=\frac{1}{n_g^2}\int dM\langle N_T(N_T-1)\rangle u(k,M)^pn_h(M) \, ,
\end{eqnarray}
where $u(k,M)$ is the NFW profile in Fourier space and $n_g$ is the galaxy number density
\begin{eqnarray}
n_g=\int dM\langle N_g(M)\rangle n_h(M) \, .
\end{eqnarray}

The second moment of the HOD that appear in eq.~\ref{1h} can be simplified as
\begin{eqnarray}
\langle N_T(N_T-1)\rangle\simeq\langle N_T\rangle^2-\langle N_h\rangle^2\, ,
\end{eqnarray}
and the power index $p$ for the NFW profile is $p=1$ when  $\langle N_T(N_T-1)\rangle<1$ and $p=2$ otherwise  \citep{Lee:2009}.

The two-halo term  of galaxy power spectrum is
\begin{eqnarray}
P^{2h}(k)&=&\left[\frac{1}{n_g}\int dM\langle N_T(M)\rangle u(k,M)n_h(M)b(M)\right]^2 \nonumber\\
&&\times P_{\rm lin}(k)\, ,
\end{eqnarray}
where $P_{\rm lin}(k)$ is the linear power spectrum and $b(M)$ is the linear bias factor calculated as in \citep{Cooray2002}. The total galaxy power spectrum is then $P_g(k)=P^{1h}(k)+P^{2h}(k)$.

The observed angular power spectrum can be related to the three-dimensional galaxy power spectrum through a redshift integration along the line of sight \citep{Knox2001,Amblard2007}:
\begin{eqnarray}
C_{\ell}^{\nu \nu'}=\int dz\left(\frac{d\chi}{dz}\right)\left(\frac{a}{\chi}\right)^2\bar{j}_{\nu}(z)\bar{j}_{\nu'}(z)P_g(\ell/\chi,z) \, ,
\end{eqnarray}
where $\chi$ is the comoving radial distance, $a$ is the scale factor and $\bar{j}_{\nu}(z)$ is the mean emissivity at the
frequency $\nu$ and redshift $z$ per comoving unit volume that can be obtained as:
\begin{eqnarray}\label{jnu}
\bar{j}_{\nu}(z)=\int dL\phi(L,z)\frac{L}{4\pi} \, .
\end{eqnarray}
Here the luminosity function is
\begin{eqnarray}
\phi(L,z)dL=dL\int dMP(L|M)n_T(M,z) \, .
\end{eqnarray}

To fit data at different frequencies we assume that the luminosity-mass
relation in the IR follows the spectral energy distribution (SED) of a modified
black-body (here we normalize at $250\mu m$ at $z=0$) with
\begin{equation}
L_\nu(M)=L_{250}(M)(1-e^{-\tau})B(\nu_0,T_d)\, ,
\end{equation}
where $T_d$ is the dust temperature and the optical depth is $\tau=\left(\frac{\nu_0}{\nu}\right)^{\beta_d}$ when
$B(\nu_0,T_d)$ is the Planck function and $L_{250}(M)$ is given by eq~(\ref{lm_relation}).

The final power spectrum is a combination of galaxy clustering, shot-noise and the Galactic cirrus such that
$C_l^{\rm tot} = C_l^{\rm CFIRB} + C_l^{\rm cirrus} + C_l^{\rm SN}$, where $C_l^{\rm CFIRB}$ is the power spectrum
derived above and $C_l^{\rm SN}$ is the scale-independent shot-noise. 
To account for the Galactic cirrus contribution to the CFIRB, we add to the predicted angular
power spectrum a cirrus power-law power spectrum with the same shape of that
used by \citet{Amblard:2011gc}, where the authors assumed the same cirrus
power-law power-spectrum shape from measurements of IRAS and MIPS \citep{Lagache2007} at
$100$\,$\mu$m with  $C_l \propto l^{-n}$ with $n=2.89\pm0.22$.  In Amblard et al. (2011) this 100\,$\mu$m spectrum was
extended to longer wavelengths using the spectral dependence of \citet{schlegel}.
Here we rescale the amplitude of the cirrus power spectrum with amplitudes
 $C^{\rm cirrus}_{i}$ at each of the three wavelengths ($i=250,350$ and 500 $\mu$) taken to be free parameters and model-fit those three parameters describing the amplitude
 as part of the global halo model fits with the MCMC approach.

\begin{figure}
    \begin{center}
        \includegraphics[scale=0.5,clip]{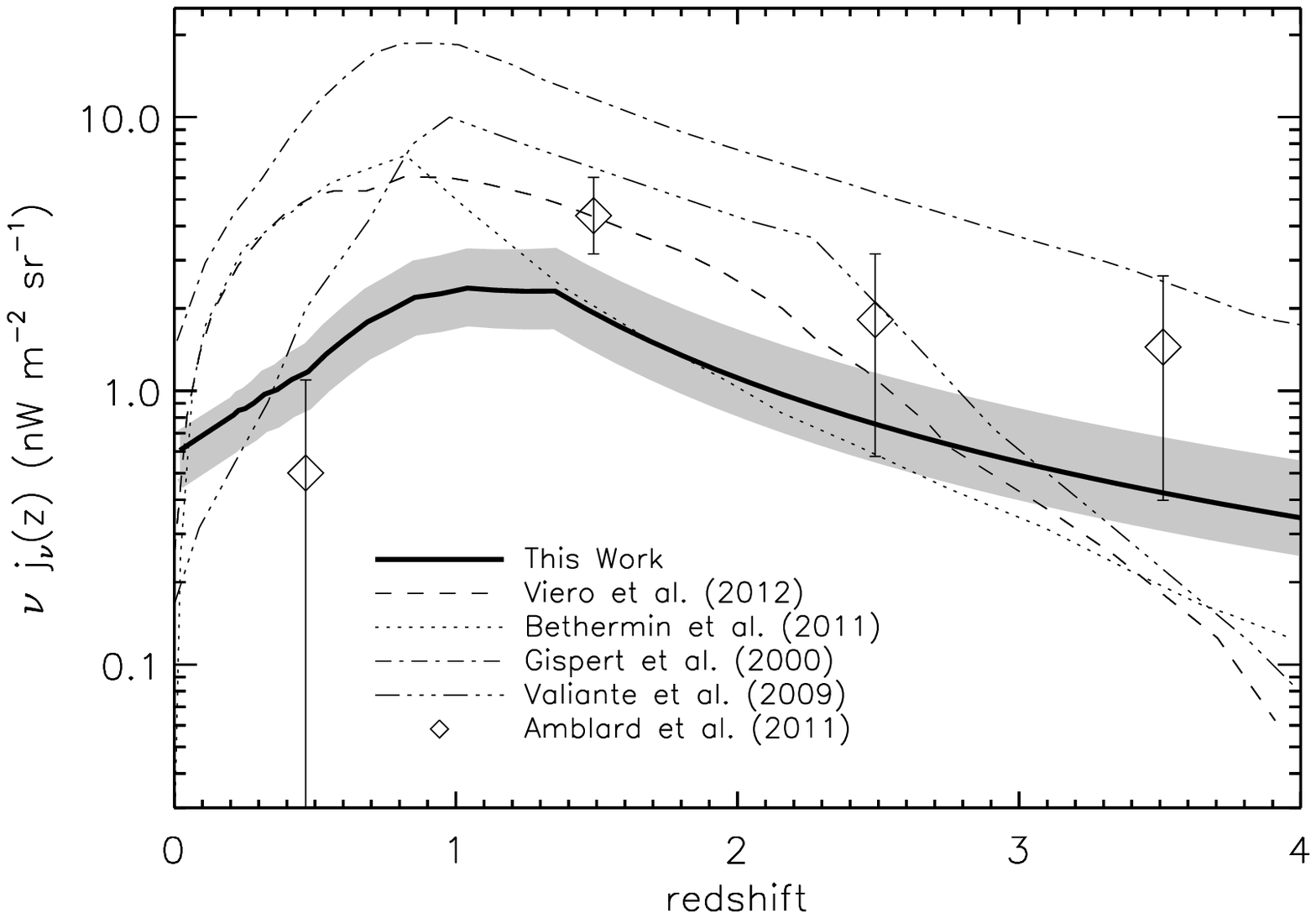}
            \caption{Best-fit determination of mean emissivity at 250\,$\mu$m as a function of the
            redshift, $\nu j_\nu(z)$ (thick solid line), and its 1$\sigma$ error from the MCMC model fits (grey shaded region)
for sources with $S_{250} < 50$mJy.
 We show several model predictions from the
            literature \citep{Valiante2009,Bethermin2011} and compare our
            estimates to the determinations from the halo model fits to the
            CFIRB power spectra by            \citet{Amblard:2011gc} and \citet{Viero2012}.
            \citet{Amblard:2011gc}
            measurements involve a binned
            description of $j_\nu(z)$ with 1$\sigma$ errors determined from the
            fit,
            while \citet{Viero2012} result is the best-fit relation for their work.
}
    \end{center}
    \label{fig:jnu}
\end{figure}

\section{Results and Discussion}
\label{sec:results}

We fit the halo-model described above to the $250$, $350$
and $500$\,$\mu$m CFIRB angular power spectrum data for the H-ATLAS GAMA-15 field by varying the halo model parameters
and the SED parameters. The dimension of the parameter space is thus $12$ with free parameters involving $T_d$, $\beta_d$, $\alpha_l$, $\beta_l$, $L_0$, $p_M$,  $C^{\rm cirrus,l=230}_{i}$ and $SN_{i}$. 
We make use of a Markov Chain Monte Carlo (MCMC) procedure, modified from the publicly available CosmoMC \citep{Lewis2002},
with a convergence diagnostics based on the Gelman-Rubin
criterion \citep{gelman}. To keep the number of free parameters in the halo model manageable, we a priori constrain the
 $M_{0l}$ in equation~\ref{mol_evolution} to the value of  $\log M_{0l}/M_{\sun} = 11.5 \pm 1.7$ as determined
by a fit to the low-redshift luminosity function at 250\,$\mu$m \citep{DeBernardis:2012bf} using data 
from Vaccari et al. (2010) and Dye et al. (2010). The best fit parameters and the uncertainties from the halo model fits are listed in Table \ref{tab:tab_fit}.

\begin{figure}
    \begin{center}
        \includegraphics[scale=0.35,clip]{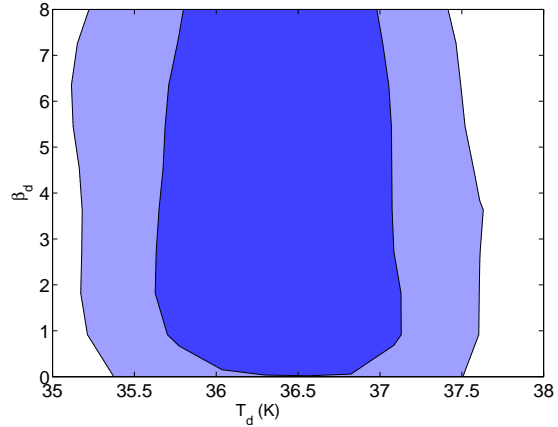}
            \caption{68\% and 95\% confidence level constraints on $T_d$ and $\beta_{\mathrm{dust}}$.}
    \end{center}
    \label{fig:td_bd}
\end{figure}  

\subsection{Cirrus Amplitude and Cirrus Dust Temperature}

We now discuss some of the results starting from our constraints on the cirrus fluctuations.
The cirrus amplitudes have values of $(3.5 \pm 1.3) \times 10^5, (1.2 \pm 1.0) \times 10^4$ and $(1.1 \pm 0.9) \times 10^3$ Jy$^2$/sr at 250, 350 and 500 $\mu$m, respectively,
at $\ell=230$ corresponding to 100 arcminute angular scales.
These values are comparable to the cirrus amplitudes in the Lockman-Hole determined by Amblard et al. (2011). The GAMA-15 area we have used for this study is thus comparabale to some of the least Galactic cirrus contaminated fields
on the sky. For comparison, the GAMA 9 hour area studied by Bracco et al. (2011)  has cirrus amplitudes of $\sim 3\times10^7,2\times 10^6$ and $1\times 10^5$ at 250, 350 and 500 $\mu$, respectively. These are roughly
a factor of 100 larger than the cirrus fluctuation amplitude in the GAMA-15 areas used here. The third field we considered for this study in GAMA 12 hour area was found to have cirrus amplitudes that are roughly
a factor of 20 to 30 larger.

In order to determine if the cirrus dust in the GAMA-15 field is comparable to dust in the high cirrus intensity regions such as the GAMA-9 field, we fitted a modified blackbody model to the cirrus rms fluctuation amplitude.
We found the dust temperature and the dust emissivity parameter $\beta$ to be $21.1 \pm 1.9$ K and $2.9 \pm 0.8$, respectively. The results from the same analysis at 100 arcminute-scale rms fluctuations
are $20.1 \pm 0.9$ and $1.3 \pm 0.2$ for dust temperature and emissivity, respectively. Even though the cirrus amplitude is lower with rms fluctuations, $\sqrt{C_{i}^{\rm cirrus}}$, at a factor of 10 below
the GAMA-9 area studied in Bracco et al. (2011), we find the dust temperature to be comparable. It is unclear if the difference in the dust emissivity parameter is significant or captures any physical variations in the dust
from high to low cirrus intensity, especially given the well-known degeneracy between dust temperature and $\beta$.  Fluctuation measurements in 
all of the 600 sq. degree H-ATLAS fields should allow a measurement of $\beta$ as a function of cirrus amplitude.

\begin{figure*}
    \begin{center}
    \includegraphics[scale=0.85,clip]{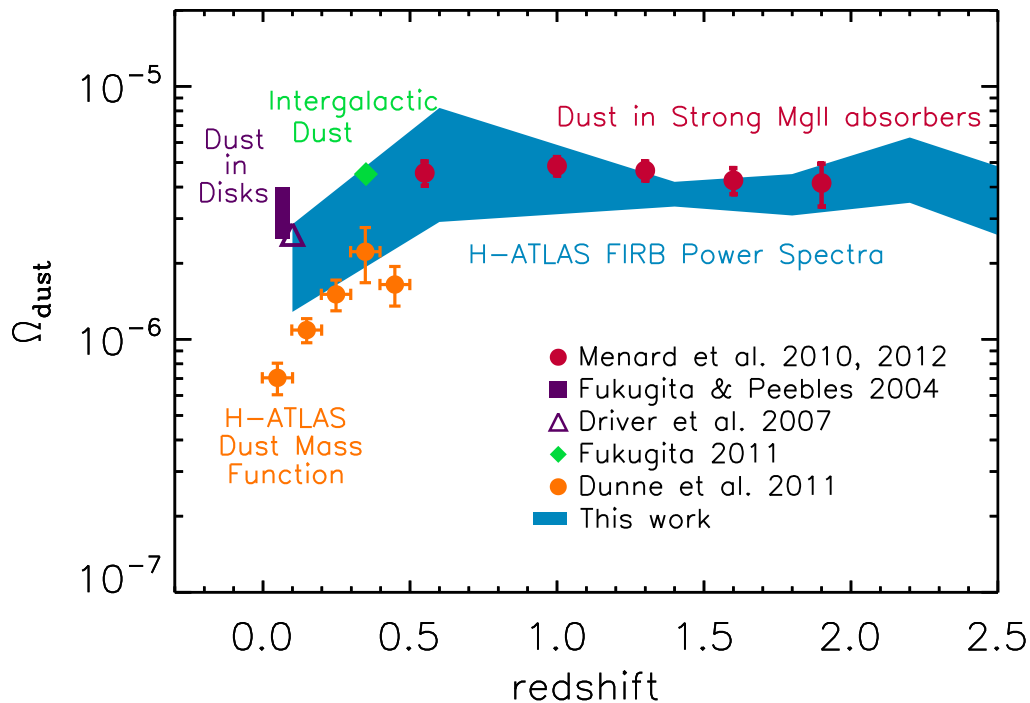}
    \end{center}
    \caption{The cosmic density of dust $\Omega_{dust}$ vs redshift as determined from the CFIRB
    power spectra from H-ATLAS GAMA-15 field (shaded region). The thickness of the region
    corresponds to the $1-\sigma$ ranges of  the halo model parameter uncertainties as determined by
    MCMC fits to the data (Table~1). We also compare our estimate to previous measurements in the
    literature. The measurements labeled H-ATLAS dust mass function are from the low-redshift dust mass function measurements in Dunne et al. (2011).
The other estimates are based on extinction measurements from the SDSS (e.g., M\'enard et al. 2010, 2012; Fukugita 2011; Fukugita \& Peebles 2004) and 2dF (e.g., Driver et al. 2007).
}
\label{fig:dust_plot}
\end{figure*}

\subsection{Faint Star-Forming Galaxy Statistics}

Moving to the galaxy distribution, in Fig.~7, we show the halo occupation distribution at $z=1$
corresponding  to the best-fit values of the parameters and the $1\sigma$
uncertainty region for three different luminosity cut-off values. 
At $z=1$, as show in Fig.~7, for $L_{\rm IR}>10^9$ L$_{\sun}$ galaxies, the HOD drops quickly for masses smaller than $\log(M_{\rm
min}/M_{\odot})\simeq10.7$ and the high-mass end has a power-law behavior with
a slope $\sim1$.  By design, this halo model based on CLFs has the advantage that it does not lead to unphysical situations with power-law
slopes for the HOD greater than one as found by Amblard et al. (2011).

Both the HOD and the underlying luminosity-mass relations are consistent
with \citet{DeBernardis:2012bf}, where a similar model was used to reinterpret Amblard et al. (2011) anisotropy measurement. The key difference between the
work of \citet{DeBernardis:2012bf} and the work here is that we introduce a dust SED to model-fit power spectra measurements in the three wavebands of SPIRE, while
in earlier work only 250\,$\mu$m measurements were used for the model fit. For comparison with recent model descriptions of the CFIRB power spectrum, we also calculate
the effective halo mass scale given by
\begin{equation}
M_{\rm eff}= \int dM n_h(M) M \frac{N_T(M)}{n_g} \, .
\end{equation}
With this definition we find $M_{\rm eff}=3.2\times10^{12}$ M$_{\rm sun}$ at $z=2$, consistent with the effective mass scales of
\citet{Shang:2011mh} and \citet{DeBernardis:2012bf}  of $M_{\rm eff}\sim4\times10^{12}$ and slightly lower than the value of $\sim5\times10^{12}$
from \citet{Xia2012}.

The MCMC fits to the CFIRB power spectrum data show that the charicteristic mass scale $M_{0l}$ evolves with
redshift as $(1+z)^{-2.9 \pm 0.4}$. In order to compare this with existing models, we convert this evolution in the charachteristic mass scale to an evolution of the
$L(M,z)$ relation. As $L(M) \propto (M/M_{0l})^{-\alpha_l}$, we find $L(M,z) \propto M^\alpha_l (1+z)^{-p_M\alpha_l}$. Using the best-fit values, we find
 $L(M,z) \propto M^{0.70 \pm 0.05}(1+z)^{2.0\pm0.4}$. In Lapi et al. (2011), their equation 9 with the SFR as a measure of the IR luminosity, this relation is
expected to be $M(1+z)^{2.1}$. In Dekel et al. (2009), the expectation is $M^{1.15}(1+z)^{2.25}$. While we find a lower value for the power-law dependence on the halo mass with IR luminosity,
the redshift evolution is consistent with both these models. 

To test the overall consistency of our model relative to existing observations at the bright-end, in Fig.~8, we compare the predicted luminosity functions 250 $\mu$m-selected galaxies
in several redshift bins with existing measurements in the literature from Eales et al. (2010) and Lapi et al. (2011).  The former relies on the spectroscopic redshifts in GOODS fields while the latter makes
use of photometric redshifts. We find the overall agreement to be adequate given the uncertainties in the angular power spectrum and the resulting parameter uncertainties of the halo model.
In future, the overall modeling could be improved with a joint-fit to both the angular power spectra and the measured luminosity functions.

In Fig.~9 we show the predicted redshift distributions of the 250 $\mu$m-selected galaxies in two 250-$\mu$m flux density bins in our model with a comparison to a measured redshift
distribution with close to 900 optical spectra of Herschel-selected galaxies with Keck/LRIS and DEIMOS in Casey et al. (2012).
While there is an overall agreement, for the brighter flux density bin, the measured redshift distribution shows a distinct tail a small, but non-negligible, fraction of  galaxies
at $ z > 2$. It is unclear if those redshifts suggest the presence of bright galaxies that are lacking in our halo model or if those redshifts are associated with 
lensed sub-mm galaxies \citep{Negrello2010, Wardlow2012} with intrinstic fluxes that are below 10 mJy. If lensed, due to magnification boost, such fainter galaxies will appear in the brighter bin. 
We also note that the current halo model ignores any lensing effect in the anisotropy power spectrum. Existing models suggest that the lensing rate at 250 $\mu$m with flux
densities below 50 mJy is small. At 500 $\mu$m, however, the lensed counts are at the level of 10\% (Wardlow et al. 2012). 
While we do not have the signal-to-noise ratio for a lensing analysis of the far-IR background anisotropies with the current data and the power spectrum, a future goal of sub-mm anisotropy
studies must involve characterizing the lensing modification to the power spectrum.

In Fig.~10 we show the redshift evolution of the emissivity predicted by our model at 250 $\mu$m according to equation \ref{jnu}. The shaded region shows 
the $1\sigma$ uncertainty associated with the best-fit model. For comparison we show the results of
\citet{Viero2012, Valiante2009,Bethermin2011,Amblard:2011gc, Gispert2000}. The distribution predicted
by our fit is consistent for a wide range of redshifts (up to $z>3$)
with \citet{Viero2012, Valiante2009,Bethermin2011,Amblard:2011gc}. The
recent fit of \citet{Viero2012} to the HerMES angular power spectra shows a lower emissivity at both low-redshift ($z < 0.5$) and high-redshift ($z>2.5$) ends.

The excess in the emissivity at the low-redshifts ($z < 0.1$)  partly explains the difference in the power spectrum amplitude at 250 $\mu$m
between the previous angular power spectra and H-ATLAS data. As discussed earlier, the GAMA-15 field of H-ATLAS is known to contain an overdensity
of low-redshift galaxies. The brightest of these sources with $S_{250} > 50$ mJy are clearly visible in Fig.~2 when comparing the original and masked maps. While the
$S_{250} > 50$ mJy  mask is expected to remove a substantial fraction of the low-z population, we expect a fraction of the fainter  ones to remain. Such galaxies
are not present in the well-known extragalactic fields of the HerMES survey such as Lockman-Hole and the NDWFS-Bootes field. The difference is a factor of $\sim$ 2 amplitude increase
in the power spectrum at 250 $\mu$m. As the excess population is primarily at low redshifts, the difference only shows up at 250 $\mu$m, while we do not see any
significant difference at 350 and 500 $\mu$m between HerMES and H-ATLAS power spectra. The final result of this is to increase the emissivity at 250 $\mu$m at lowest redshifts $ z< 0.1$ in our model
relative to the emissivity function derived in \citet{Viero2012}. 

The H-ATLAS GAMA-15 field also shows an overall increase of bright counts at 250\,$\mu$m relative to the HerMES fields (Rigby et al. in prep) and
we verified that our suggestion of a factor of 2 increase in the power spectrum is coming from low-redshift galaxies is consistent with the differences in the number counts.
The difference in the counts also explains the increase in the shot-noise at 250 $\mu$m relative to the value found in HerMES power spectra. These differences generally 
suggest that large field-to-field variations in the angular power spectrum with variations well above the typical Gaussian cosmic variance calculations. Such variations are readily visible when comparing
individual field power spectra in \citet{Viero2012} (their Fig.~3). 

Through our joint model-fit to 250, 350 and 500 $\mu$m power spectra, we also determine the SED of far-IR background anisotropies. To keep the number of free parameters in our model
small, here we assume that the  SED can be described by an isothermal black-body model. The best-fit dust temperature value that describes the far-IR fluctuations is $37 \pm 2$ K
while the emissivity parameter $\beta$ is unconstrained. In Fig.~11 we show the best-fit 68\% and 95\% confidence level intrvals of $T_d$ and $\beta$, after marginalizing over all
other parameters of the halo model. This figure makes it clear why we are not able to determine $\beta$ with the current data due to  degeneracies between the model parameters.
The dust temperature we measure should be considered as the average dust temperature of all galaxies that is contributing to far-IR background anisotropy power spectrum.
The dust temperature is higher than the typical 20K dust temperature derived from the absolute background spectra at far-IR wavelengths from experiments such as FIRAS and Planck \citep{FIRAS}. 

This difference in the dust temperature could be understood since the absolute measurements, especially at degree angular-scale beams, are likely to be dominated by
the Galactic cirrus, and thus the temperature measurement could be biased low. The dust temperature we measure from the far-IR power spectra is fully consistent with the
the value of $44 \pm 7$K by Shang et al. (2011) in their modeling of the Planck far-IR power spectra (assuming the fixed value $\beta=2$). Thus, while the best-fit SED model
of the absolute CIB may suggest a low temperature value, the anisotropies from approximately 1-30 arcminute angular scales follow a SED with a higher dust temperature value.
Separately, we also note that our dust temperature of 37 $\pm 2$ K is also consistent with the average dust temperature values of $36 \pm 7$ K 
\citep{Chapman2005, Dunne:2000xd} for high-z SCUBA-selected sub-mm galaxies, but is somewhat higher than the average dust temperature value of $28 \pm 8$ K
for Herschel-selected bright galaxies in \citet{Amblard2010}. The Amblard et al. (2010) value is dominated by low-redshift  ($z \sim 0.1$) galaxies with Herschel identifications
to SDSS redshifts. In the local Universe, most dusty late-type galaxies show cold dust with temperatures around 20 K \citep{Galametz2012, Davies2012}.
The higher temperature we find for the far-IR background anisotropies then suggests that the average interstellar radiation field in galaxies at $z \sim 1$ to 2 that dominate the
dust emissivity is higher by a factor of $2^6$ when compared that local late-type galaxies.

\subsection{Cosmic Dust Abundance}

The model described above allows us to estimate the fractional cosmic dust density:
\begin{eqnarray}\label{omegadust}
\Omega_{\rm dust}(z)=\frac{1}{\rho_0}\int_{L_{min}} dL\phi(L,z)M_{dust}(L) \, ,
\end{eqnarray}
where $M_{dust}$ is the dust mass for a given IR luminosity and $\rho_0$ is the
critical density of the Universe. Here we make use $\phi(L,z)$ as derived by the halo model fits to the far-IR background power spectra.

To convert luminosities to dust mass, we follow equation~4 in \citep{Fu:2012ba}. This requires an assumption related to the dust mass absorption coefficient, $\kappa_d$.
It is generally assumed that the opacity follows $\kappa_d(\nu) \propto \nu^\beta$ with a normalization of $\kappa_d=0.07 \pm 0.02$ m$^2$ kg$^{-1}$ at 850 $\mu$m
\citep{Dunne:2000xd, James2002}. This normalization, unfortunately, is highly uncertain and could easily vary by a factor of few or more
 (see discussion in James et al. 2002). The value we adopt here is appropriate for dusty galaxies and matches well with the integrated spectrum of the Milky Way.

The conversion to dust mass also requires the SED of dust emission. Here we make use of the average dust temperature value of $37 \pm 2$ K as determined by the model-fits
to the angular power spectra. As $\beta$ is undetermined from the data, we take its range with a prior between 1 and 2.5, consistent with typical values of 1.5 or 2 that is
generally assumed in the literature. When calculating  $\Omega_{\rm dust}(z)$ we marginalize over all parameter uncertainties so that we fully capture the full likelihood from
 the MCMC chains given the prior on $\beta$. Note that our assumption of a constant dust temperature is at odds with local late-type spirals that show much lower temperatures.
However, it is also known that there are some sub-mm galaxies, especially those that are radio bright, with dust temperatures in the excess of 60 K. Thus with a value of 37 $\pm 2$ K
we may be using a representative average value for the dust temperature and an average of the dust SED for all galaxies at a variety of redshifts. Finally there are some indications that
the dust temperature is IR luminosity dependent (see the discussion in Amblard et al. 2010). If that remains to be the case then the correct approch with eq.~\ref{omegadust} will be
to take into account that luminosity dependence as seen in the observations. Given that the current indications are coming from small galaxy samples, we do not pursue such a correction, but
highlight that future studies could improve our dust abundance estimate.

In Fig.~12 we show our results. In addition to the direct emission estimate that we have considered here, we also show the low-z dust abundances by integrating over the dust mass functions
in Dunne et al. (2011). Those dust mass functions are limited to $z < 0.5$ due to the limited availability of spectroscopic data at higher redshifts. While in principle
dust mass function captures the total dust of detected galaxies, the mass functions can be extrapolated to the faint-end, as has been done here, to acocunt for the fainter
populations below individual detection levels. Thus the abundances from mass functions must agree with the estimates based on the anisiotropy measurements. We do not use our halo model
to estimate the dust abundance at $z < 0.05$ since our halo model is  normalized to the luminosity function of dusty galaxies at low redshifts.

Our measurement  indicates that the dust density ranges between $\Omega_{\rm dust}\simeq10^{-6}$ to
$8\times10^{-6}$ in the redshift range $z=0.5-3$. We note that  the $\Omega_{\rm
dust}$ prediction of this work has a smaller uncertainty than that  in
\citet{DeBernardis:2012bf} where the estimation was done assuming a larger
range for $T_d$ and $\beta_d$. In equation~\ref{omegadust} we integrate over luminosities $L > 10^9$ L$_{\sun}$. However 
in this calculation the choice of minimum luminosity is less relevant, since the uncertainty on the dust density estimate is
dominated by the large uncertainties of dust temperature and spectral emissivity index $\beta$.

Fig.~\ref{fig:dust_plot}  also summarizes the dust-density measurements of
\citet{Fukugita2004, Driver2007, Menard2010, Fukugita2011}. 
We have combined the points from Menard et al. (2010) for the
dust contributions of halos and those from
Menard \& Fukugita, (2012) to a single set of data points, under the assumption that the
amount of dust in halos doesn't evolve significantly with redshift. This is an assumption and could be tested in future data.
At high redshifts our estimate is consistent with the results of \citet{Fukugita2004, Driver2007, Menard2010}.

We note that the M\'enard et al. (2010, 2012) measurements assume a reddening law appropriate for the SMC. A Milky Way reddening law would have
resulted in a a factor of 1.8 higher dust masses, and thus dust abundance, than the values shown in Fig.~12 (see discussion in M\'enard et al. 2012).
Since the extinction measurements make use of a reddening law consistent with SMC while the direct emission measurements of the dust abundance that we show
here assumed a dust mass absorption coefficient that is more consistent with the Milky Way, it is interesting to ask why the two measurements shown in Fig.~12 agree.
The UV reddening law related to extinction based dust abundance estimates comes from small grains that dominate the absorption and scattering surface area. SMC differs
from other galaxies in that it does not show a prominent 2200$\AA$ feature, which is assumed to come from carbon bonds.
On the other hand, the far-IR emission that we have detected is likely dominated by large grains, usually assumed to be a mixture of silicates and carbonaceous grains.
The difference in the reddening law between SMC and Milky Way then should not complicate the abundance estimates since extinction and emission may be
coming from different populations of dust grains (e.g., \citet{LiDraine2002}). 

While  Fig.~12 is showing that the dust abundances from extinction measurements are consistent with direct emission measure from far-IR background fluctuations, the above discussion
may suggest that this comparison is incomplete. It could be that this agreement is merely a concidence of two different populations. Thus the total abundance of the
dust in the Universe is likely at most the total when summing up extinction and emission measurements. However, a direct summation of the two measurements is misleading
and likely leads to an overestimate. While small and large grains dominate extinction and emission, respectively, the two effects are not exclusive in terms of the different
populations of dust grains. Some of the grains associated with extinction must also be responsible for emission. 

The far-IR background anisotropy measurements we have presented here
have the advantage they capture the full population of grains responsible for thermal dust emission in galaxies. The extinction measurements, however, are biased
to clean lines of sights where the lines of sights do not cross the galactic disks. We have corrected for the missing dust in disks by adding the density of dust in disks
at $z \sim 0.3$ to all measurements at high redshifts, but the disk dust density could easily evolve with redshift. The agreement we find here between the two different sets of
measurements may, however, argue that there is no significant evolution in the dust density in galactic disks. In any case we suggest that one does not
derive quick conclusions on the dust abundances or the agreements between extinction and emission measurements as shown in Fig.~12. There are built-in assumptions
and biases between different sets of measurements and future studies must improve on the current analyses to understand the extent to which extinction and emission
measurements can be used to obtain the total dust content of the universe.

While the {\it Herschel} fluctuation measurements
have the advantage we see total emission, they have the disadvantage that we cannot separate the dust in disks to diffuse dust in halos that should also be emitting
at far-IR wavelengths. In future, it may be possible to separate the two based on cross-correlation studies of far-IR fluctuations with galaxy catalogs and using stacking
analysis, especially for galaxy populations at low redshifts. These are some of the studies that we aim to explore with the H-ATLAS maps in upcoming papers.

%\appendix

\begin{table*}
\caption{Angular power spectrum measurements at 250, 350 and 500\,$\mu$m from GAMA-15 field of H-ATLAS. We tabulate the values as $l^2C_l/2\pi$ without shot-noise subtracted.} 
\begin{center}
\begin{tabular}{| c  c | c  c | c c | }
\hline
\multicolumn{2}{|c|}{250\,$\mu$m}&
\multicolumn{2}{|c|}{350\,$\mu$m}&
\multicolumn{2}{|c|}{500\,$\mu$m}\\
\hline
$l $&$ l^2C_l/2\pi \: [\mathrm{Jy}^2/\mathrm{Sr}^2] $&$ l $&$ l^2C_l/2\pi \: [\mathrm{Jy}^2/\mathrm{Sr}^2] $&$ l $&$ l^2C_l/2\pi \: [\mathrm{Jy}^2/\mathrm{Sr}^2]$\\
\hline
$ 2.30 \times10^2$ & $(2.33 \pm 1.49) \times10^{10}  $&$ 2.45\times10^2 $&$ (2.71 \pm 1.65) \times10^{9}  $&$  1.58\times10^2  $&$ (5.28 \pm 5.09) \times10^{8}  $\\   
$ 2.94 \times10^2$ & $(1.78 \pm 0.89) \times10^{10}  $&$ 3.11\times10^2 $&$ (2.41 \pm 1.15) \times10^{9}  $&$  1.99\times10^2  $&$ (1.53 \pm 1.17) \times10^{8}  $\\  
$ 3.76 \times10^2$ & $(1.07 \pm 0.42) \times10^{10}  $&$ 3.95\times10^2 $&$ (2.37 \pm 0.90) \times10^{9}  $&$  2.52\times10^2  $&$ (3.62 \pm 2.18) \times10^{8}  $\\  
$ 4.80 \times10^2$ & $(8.75 \pm 2.70) \times10^{9}   $&$ 5.02\times10^2 $&$ (2.09 \pm 0.62) \times10^{9}  $&$  3.18\times10^2  $&$ (3.71 \pm 1.78) \times10^{8}  $\\  
$ 6.14 \times10^2$ & $(1.17 \pm 0.28) \times10^{10}  $&$ 6.38\times10^2 $&$ (4.87 \pm 1.14) \times10^{9}  $&$  4.02\times10^2  $&$ (8.47 \pm 3.22) \times10^{8}  $\\  
$ 7.85 \times10^2$ & $(6.96 \pm 1.33) \times10^{9}   $&$ 8.11\times10^2 $&$ (2.79 \pm 0.52) \times10^{9}  $&$  5.07\times10^2  $&$ (5.88 \pm 1.78) \times10^{8}  $\\  
$ 1.00 \times10^3$ & $(8.05 \pm 1.20) \times10^{9}   $&$ 1.03\times10^3 $&$ (3.81 \pm 0.56) \times10^{9}  $&$  6.41\times10^2  $&$ (8.29 \pm 1.99) \times10^{8}  $\\  
$ 1.28 \times10^3$ & $(7.80 \pm 0.91) \times10^{9}   $&$ 1.31\times10^3 $&$ (3.68 \pm 0.43) \times10^{9}  $&$  8.09\times10^2  $&$ (9.33 \pm 1.78) \times10^{8}  $\\  
$ 1.64 \times10^3$ & $(1.32 \pm 0.12) \times10^{10}  $&$ 1.67\times10^3 $&$ (6.81 \pm 0.63) \times10^{9}  $&$  1.02\times10^3  $&$ (1.43 \pm 0.22) \times10^{9}  $\\  
$ 2.10 \times10^3$ & $(1.43 \pm 0.11) \times10^{10}  $&$ 2.12\times10^3 $&$ (7.53 \pm 0.57) \times10^{9}  $&$  1.29\times10^3  $&$ (1.60 \pm 0.19) \times10^{9}  $\\  
$ 2.68 \times10^3$ & $(1.63 \pm 0.10) \times10^{10}  $&$ 2.69\times10^3 $&$ (9.28 \pm 0.57) \times10^{9}  $&$  1.63\times10^3  $&$ (2.46 \pm 0.24) \times10^{9}  $\\  
$ 3.42 \times10^3$ & $(2.45 \pm 0.12) \times10^{10}  $&$ 3.42\times10^3 $&$ (1.39 \pm 0.07) \times10^{10} $&$  2.06\times10^3  $&$ (3.21 \pm 0.26) \times10^{9}  $\\  
$ 4.38 \times10^3$ & $(3.33 \pm 0.14) \times10^{10}  $&$ 4.35\times10^3 $&$ (2.01 \pm 0.09) \times10^{10} $&$  2.60\times10^3  $&$ (4.34 \pm 0.29) \times10^{9}  $\\   
$ 5.59 \times10^3$ & $(5.04 \pm 0.19) \times10^{10}  $&$ 5.52\times10^3 $&$ (3.04 \pm 0.12) \times10^{10} $&$  3.29\times10^3  $&$ (5.76 \pm 0.34) \times10^{9}  $\\   
$ 7.15 \times10^3$ & $(7.14 \pm 0.24) \times10^{10}  $&$ 7.02\times10^3 $&$ (4.30 \pm 0.15) \times10^{10} $&$  4.15\times10^3  $&$ (8.05 \pm 0.43) \times10^{9}  $\\   
$ 9.14 \times10^3$ & $(1.11 \pm 0.04) \times10^{11}  $&$ 8.92\times10^3 $&$ (6.44 \pm 0.22) \times10^{10} $&$  5.24\times10^3  $&$ (1.21 \pm 0.06) \times10^{10} $\\   
$ 1.17 \times10^4$ & $(1.69 \pm 0.05) \times10^{11}  $&$ 1.13\times10^4 $&$ (9.92 \pm 0.33) \times10^{10} $&$  6.62\times10^3  $&$ (1.72 \pm 0.08) \times10^{10} $\\    
$ 1.49 \times10^4$ & $(2.54 \pm 0.07) \times10^{11}  $&$ 1.44\times10^4 $&$ (1.53 \pm 0.05) \times10^{11} $&$  8.36\times10^3  $&$ (2.53 \pm 0.12) \times10^{10} $\\         
$ 1.91 \times10^4$ & $(4.01 \pm 0.12) \times10^{11}  $&$ 1.83\times10^4 $&$ (2.44 \pm 0.08) \times10^{11} $&$  1.06\times10^4  $&$ (3.71 \pm 0.18) \times10^{10} $\\         
$    \           $ & $  \                            $&$ 2.33\times10^4 $&$ (3.92 \pm 0.12) \times10^{11} $&$  1.33\times10^4  $&$ (5.79 \pm 0.29) \times10^{10} $\\         
$    \           $ & $  \                            $&$   \            $&$  \                            $&$  1.69\times10^4  $&$ (9.04 \pm 0.45) \times10^{10} $\\         
$    \           $ & $  \                            $&$   \            $&$  \                            $&$  2.13\times10^4  $&$ (1.55 \pm 0.07) \times10^{11} $\\         
\hline                                                                                                                 
\end{tabular}         
\end{center}          
\label{tab:power_data}
\end{table*}

\section{Conclusions}
\label{sec:conclusion}

We have analyzed the anisotropies of the cosmic far-infrared background in the
GAMA-15 Herschel-ATLAS field using the SPIRE data in the $250$, $350$ and $500$\,$\mu$m
bands. The power spectra are found to be consistent with previous estimates, but with a higher
amplitude of clustering at 250\,$\mu$m. We find this increase in the amplitude
and the associated  increase in the shot-noise to be coming from an increase in the surface density
of low-redshift galaxies that peak at 250\,$\mu$m. The increase is
also visible in terms of the bright source counts of the H-ATLAS GAMA fields (e.g., Ribgy et al. in prep).

We have used a conditional luminosity function approach to model the anisotropy power spectrum of the far-infrared background. In order to fit H-ATLAS power spectra at
the three wavebands of SPIRE we have adopted the spectral energy distribution of a
modified black body and constrained the dust parameters $T_d$ and $\beta_d$ using a joint fit to
power spectra at 250, 350 and 500\,$\mu$m.
The results of our fit substantially confirm previous results from the analysis
of {\it Herschel} data and allow us to improve the constraints on the cosmic dust
density that resides in the star forming galaxies responsible for the
far-infrared background. We have found that the fraction of dust with respect
to the total density of the Universe is $\Omega_{dust}=10^{-6}$ to
$8\times10^{-6}$, consistent with estimations from observations of reddening of
metal-line absorbers.

\acknowledgments

We thank Brice M\'enard and Marco Viero for useful discussions.
The {\it Herschel}-ATLAS is a project with Herschel, which is an ESA space observatory with science instruments provided by European-led Principal Investigator consortia 
and with important participation from NASA. The H-ATLAS website is http://www.h-atlas.org/.
This work was supported by NSF CAREER AST-0645427 and NASA NNX10AD42G at UCI to AC, 
and support for US Participants in {\it Herschel} programs from NASA Herschel Science Center/JPL.


\begin{thebibliography}{99}


\bibitem[Amblard \& Cooray, 2007]{Amblard2007}
Amblard, A. \& Cooray, A. 2007, ApJ, 670, 903

\bibitem[Amblard et al., 2010]{Amblard2010}
Amblard, A., Cooray, A., Serra, P., et al. 2010, A\&A, 518, L9+

\bibitem[Amblard et al., 2011]{Amblard:2011gc}
Amblard, A., et al. 2011, Nature, 470, 510

\bibitem[Bacon et al., 2000]{BaconRE}
Bacon, D., Refregier, A., \& Ellis, R. 2000, MNRAS 318, 625

\bibitem[Berta et al., 2011]{Berta2011}
Berta, S., et al.  2011, A\&A 532, A49

\bibitem[Bethermin et al., 2011]{Bethermin2011}
Bethermin, M., Dole, H., Lagache, G., et al. 2011, A\&A, 529, A4

\bibitem[Bracco et al., 2010]{Bracco:2010ak}
Bracco, A., Cooray, A.~, Veneziani, M.~, et al. 2011, MNRAS, 412, 1151

\bibitem[van de Bosch et al., 2005]{vandenBosch:2004zs}
van de Bosch, F.C., Tormen, G., Giocoli, C. 2005, MNR

\bibitem[Cantalupo et al, 2009]{Cantalupo:2009if}
  Cantalupo C.~M.~, Borrill J.~D.~, Jaffe, A.~H.~, et al. 2009 Astrophys.\ J.\ Suppl.\, 187 (2010) 212

\bibitem[Casey et al., 2012]{Casey2012}
Casey, C. M., Berta, S., Bethermin, M., et al. 2012, ApJ accepted

\bibitem[Chapman et al., 2005]{Chapman2005}
Chapman, J.~F.~ \& Wardle, M.~ 2006 MNRAS, 371, 513
%
\bibitem[Clements et al., 2010]{Clements2010}
Clements, D.L., Dunne, L., Eales, S. 2010, MNRAS, 403, 274

\bibitem[Cooray \& Sheth, 2002]{Cooray2002}
 Cooray, A., Sheth, R.K. 2002, PR,  372, 1

\bibitem[Cooray et al., 2010]{Cooray10}
Cooray, A.  et al. 2010, A\&A, 518, L22 

\bibitem[Cooray et al., 2012]{Cooray2012}
Cooray, A., et al., 2012, Nature, 490, 514-516

\bibitem[Coppin et al., 2006]{Coppin2006}
Coppin, K., et al. 2006, MNRAS, 372, 1621

\bibitem[Coppin et al., 2011]{Coppin:2011dp}
  Coppin, K.~E.~K.~, et al. 2011 arXiv:1105.3199 [astro-ph.CO].

\bibitem[Davies et al., 2012]{Davies2012}
Davies, J.~I.~, et al. 2012 arxiv:1210.4448 [astro-ph.CO]

\bibitem[De Bernardis \& Cooray, 2012]{DeBernardis:2012bf}
De Bernardis, F., Cooray, A.~ 2012, in press on ApJ


\bibitem[Dowell et al., 2010]{dow10}
Dowell, C.D., et al. 2010, Proc. SPIE 7731, 773136

\bibitem[Driver et al., 2007]{Driver2007}
Driver, S. P., et al. 2007, MNRAS, 379, 1022


\bibitem[Dunne et al., 2000]{Dunne:2000xd}
Dunne, L.~, Eales, S.~A.~, Edmunds, M.~G.~, et al. 2000, MNRAS, 315, 115

\bibitem[Dunne et al., 2003]{Dunne:2003va}
Dunne, L.~, Eales, S.~, Ivison, R.~, et al. 2003, Nature, 424, 285

\bibitem[Dunne et al., 2010]{Dunne:2010ka}
Dunne, L.~, Gomez, H.~, da Cunha, E.~ S., et al. 2010 MNRAS, 417, 1510



\bibitem[Dwek et al., 1998]{Dwek1998}
Dwek, E., et al. 1998, ApJ, 508, 106


\bibitem[Dye et al., 2010]{Dye2010}
Dye, S., Dunne, L., Eales, S. et al. 2010, A\&A, 518, L10

\bibitem[Eales et al., 2010]{Eales:2010vw}
 Eales, S., et al. 2010,  A\&A, 518, L23

\bibitem[Fixen et al., 1998]{Fixen1998}
Fixsen, D. J., et al. 1998, ApJ, 508, 123

\bibitem[Fu et al., 2012]{Fu:2012ba}
Fu, H., et al. 2012, arXiv:1202.1829

\bibitem[Fukugita \& Peebles, 2004]{Fukugita2004}
Fukugita, M., Peebles P. J. E., 2004, ApJ, 616, 643

\bibitem[Fukugita et al., 2011]{Fukugita2011}
Fukugita, M., 2011, arXiv, arXiv:1103.4191

\bibitem[Galametz et al., 2012]{Galametz2012}
Galametz, M.~ et al. 2012, MNRAS, 425, 763

\bibitem[Gelman \& Rubin, 1992]{gelman}
Gelman, A. \& Rubin, D. 1992, Statistical Science, 7, 457-472.                                             

\bibitem[Giavalisco \& Dickinson, 2001]{Giavalisco:2000xs}
Giavalisco, M., \& Dickinson, M. 2001, ApJ, 550, 177

\bibitem[Gispert et al., 2000]{Gispert2000}
Gistpert, R., Lagache, G., \& Puget, J.L., 2000, A\&A, 360, 1

\bibitem[Glenn et al., 2010]{Glenn2010}
Glenn, J., et al. 2010, MNRAS, 409, 109

\bibitem[Griffin et al., 2010]{Griffin:2010hp}
 Griffin, M.~J.~, Abergel, M.~J.~, Abreu, A.~, et al. 2010,  A\&A, 518, L3

\bibitem[Guo et al., 2011]{Guo11}
Guo, Q., Cole, S., Lacey, C. et al. 2011, MNRAS, arXiv.org:1011.3048

\bibitem[Hickox et al., 2010]{Hickox2012}
Hickox, R. C. et al. 2012, MNRAS, 421, 284

%\cite{Hivon:2001jp}
\bibitem[Hivon et al., 2002]{Hivon:2001jp}
  Hivon E.~, Gorski K.~M.~, Netterfield C.~B.~ et al. 2002 Astrophys.\ J.\ , 567 2                                      
  %%CITATION = ASTRO-PH/0105302;%

\bibitem[James et al., 2002]{James2002}
James, A.~, Dunne, L. et al., 2002 MNRAS, 335, 753

\bibitem[Kampen et al., 2012]{Kampen2012}
Kampen, E. van, et al. 2012, MNRAS, 426, 3455

\bibitem[Knox et al., 2001]{Knox2001}
Knox, L., et al. 2001, ApJ, 550, 7

\bibitem[Lagache et al., 2007]{Lagache2007}
Lagache, G., et al. 2007, ApJ, 665, L89

\bibitem[Lagache et al., 2000]{FIRAS}
Lagache, G., Haner, L.~M.~, et al. 2000, A\&A, 354, 247

\bibitem[Lapi et al., 2011]{Lapi:2011ca}
Lapi, A., et al. 2011, ApJ, 742, 1

\bibitem[Lee et al., 2009]{Lee:2009}
Lee {\it et al.}, ApJ 695, 368 (2009)

\bibitem[Levenson et al., 2010]{Levenson2010}
Levenson, L., Marsden, G., Zemcov, M., et al. 2010, MNRAS, 409, 83

\bibitem[Lewis \& Bridle, 2002]{Lewis2002}
Lewis, A \& Bridle, S. 2002, PRD, 66, 103511

\bibitem[Li \& Draine, 2002]{LiDraine2002}
Li, A.~, Draine, B.~T.~, 2002, Apj, 572 232

\bibitem[Maddox et al., 2010]{Maddox2010}
Maddox, S. J., et al. 2010. A\&A, 518, L11

\bibitem[Menard et al., 2010]{Menard2010}
Menard B., Scranton R., Fukugita M., Richards G., 2010, MNRAS, 405, 1025

\bibitem[Menard et al., 2012]{Menard:2012am}
Menard, B., \& Fukugita, M. 2012,  arXiv:1204.1978

\bibitem[Mitchell-Wynne et al., 2012]{MitchellWynne:2012ys}
  Mitchell-Wynne K.~, Cooray A.~, Gong Y.~ et al. 2012,  Astrophys.\ J.\,753, 23

\bibitem[Navarro et al., 1997]{NFW}
Navarro, J. F., Frenk, C. S., White, S. D. M. 1997, ApJ, 490, 493

\bibitem[Negrello et al., 2010]{Negrello2010}
Negrello M., Hopwood R., De Zotti G., et al. 2010, Science, 330, 800

\bibitem[Nguyen et al., 2010]{Nguyen2010}
Nguyen, H. T., Schulz, B., Levenson, L., et al. 2010, A\&A, 518, L5

\bibitem[Oliver et al., 2010]{hermes}
Oliver, S., et al. 2010, A\&A, 518, L21.1

\bibitem[Ott et al., 2010]{Ott:2010jz}
  Ott S.~, Centre H.~S.~ and E.~S.~Agency,
  %``The Herschel Data Processing System - HIPE and Pipelines - Up and Running Since the Start of the Mission,''
  arXiv:1011.1209 [astro-ph.IM]

\bibitem[Pascale et al., 2011]{Pascale:2011br}
  Pascale E.~, Auld R.~, Dariush A.~, et al.,arXiv:1010.5782 [astro-ph.IM]
  %``The first release of data from the Herschel ATLAS: the SPIRE images,''
  
\bibitem[Planck collaboration, 2011]{Planck2011}
  The Planck Collaboration,
%  arXiv:1101.2028 [astro-ph.CO].

\bibitem[Puget et al., 1996]{Puget1996}
Puget, J.L., Abergel, A., Bernard, J.P., et al. 1996, A\&A, 308, L5

\bibitem[Rawle et al., 2010]{Rawle:2010je}
  T.~D.~Rawle {\it et al.} 2010, A\&A ,518, L14

\bibitem[Schlegel, 1998]{schlegel}
Schlegel, D.J. 1998, ApJ, 500, 525

\bibitem[Scott et al., 2010]{Scott:2010dv}
  Scott K.~S.~, Yun M.~S.~, Wilson G.~W.~, {\it et al.}, 2010, MNRAS, 405, 2260.
  %``Deep 1.1 mm-wavelength imaging of the GOODS-S field by AzTEC/ASTE - I. Source catalogue and number counts,''
  arXiv:1003.1768 [astro-ph.CO]
  %%CITATION = ARXIV:1003.1768;%%

\bibitem[Shang et al., 2011]{Shang:2011mh}
  Shang, C., et al. 2011, arXiv:1109.1522

\bibitem[Sheth \& Tormen, 1999]{ShethTormen1999}
Sheth, R. K., Tormen, G. 1999, MNRAS, 308, 119

\bibitem[Smidt et al., 2010]{Smidt:2010ra}
  Smidt J.~, Amblard A.~, Byrnes, C.~T.~ et al. 2010 Phys.\ Rev.\  D ,81, 123007

\bibitem[Smith et al., 2011]{Smith:2011fi}
 Smith,  A.~J.~, Wang, L.~, Oliver, S.~J.~, et al.
  %``HerMES: point source catalogues from deep Herschel-SPIRE observations,''
  arXiv:1109.5186 [astro-ph.CO].

\bibitem[Vaccari et al., 2010]{Vaccari2010}
Vaccari, M., Marchetti, L., Franceschini, A., et al. 2010, A\&A, 518, L20

\bibitem[Valiante et al., 2009]{Valiante2009}
Valiante, E., Lutz, D., Sturm, E., et al. 2009, ApJ, 701, 1814


\bibitem[Viero et al., 2009]{Viero2009}
Viero, M. P., et al. 2009, ApJ, 707, 1766

\bibitem[Viero et al., 2012]{Viero2012}
Viero  M.~P.~,Wang  L.~, Zemcov M.~ et al. 2012
arXiv:1208.5049 [astro-ph.CO].

\bibitem[Wardlow et al., 2012]{Wardlow2012}
Wardlow, J.L., Cooray, A., et al., 2012, ApJ, submitted arxiv:1205.3778  

\bibitem[Xia et al., 2012]{Xia2012}
Xia, J.-Q., et al. 2012, MNRAS, 422, 1324

\end{thebibliography}
\end{document}